# Searching for Low-redshift Hot Dust-obscured Galaxies

Guodong Li[1,2,3], Jingwen Wu[1,2], Chao-Wei Tsai[1,2,4], Daniel Stern[5], Roberto J. Assef[6], Peter R. M. Eisenhardt[5], Kevin McCarthy[5], Hyunsung D. Jun[7,8], Tanio Díaz-Santos[9,10], Andrew W. Blain[11], Trystan Lambert[12], Dejene Zewdie[13,14], Román Fernández Aranda[9,15], Cuihuan Li[1], Yao Wang[16], and Zeyu Tan[17]

[1] National Astronomical Observatories, Chinese Academy of Sciences, 20A Datun Road, Beijing 100101, People's Republic of China; lgd@nao.cas.cn, jingwen@nao.cas.cn, cwtsai@nao.cas.cn
[2] University of Chinese Academy of Sciences, Beijing 100049, People's Republic of China
[3] Kavli Institute for Astronomy and Astrophysics, Peking University, Beijing 100871, People's Republic of China
[4] Institute for Frontiers in Astronomy and Astrophysics, Beijing Normal University, Beijing 102206, People's Republic of China
[5] Jet Propulsion Laboratory, California Institute of Technology, 4800 Oak Grove Drive, Pasadena, CA 91109, USA
[6] Instituto de Estudios Astrofísicos, Facultad de Ingeniería y Ciencias, Universidad Diego Portales, Av. Ejército Libertador 441, Santiago, Chile
[7] Department of Physics, Northwestern College, 101 7th St SW, Orange City, IA 51041, USA
[8] School of Physics, Korea Institute for Advanced Study, 85 Hoegiro, Dongdaemun-gu, Seoul 02455, Republic of Korea
[9] Institute of Astrophysics, Foundation for Research and Technology-Hellas (FORTH), Heraklion, GR-70013, Greece
[10] School of Sciences, European University Cyprus, Diogenes street, Engomi, 1516 Nicosia, Cyprus
[11] School of Physics and Astronomy, University of Leicester, LE1 7RH Leicester, UK
[12] ICRAR, The University of Western Australia, 35 Stirling Highway, Crawley, WA 6009, Australia
[13] Centre for Space Research, North-West University, Potchefstroom 2520, South Africa
[14] Department of Physics, College of Natural and Computational Science, Debre Berhan University (DBU), P.O. Box 445, Debre Berhan, Ethiopia
[15] Department of Physics, University of Crete, 70013, Heraklion, Greece
[16] Department of Astronomy, School of Physics and Astronomy, Shanghai Jiao Tong University, 800 Dongchuan Road, Shanghai 200240, People's Republic of China
[17] Department of Physics and Astronomy, University of Georgia, 220 Cedar St, Athens, GA 30602, USA

## Abstract

Hot dust-obscured galaxies (Hot DOGs), discovered by the "W1W2 dropout" selection at high redshifts ($z \sim 2$–4), are a rare population of hyperluminous obscured quasars. Their number density is comparable to similarly luminous type 1 quasars in the same redshift range, potentially representing a short, yet critical stage in galaxy evolution. The evolution in their number density toward low redshift, however, remains unclear as their selection function is heavily biased against objects at $z \lesssim 2$. We combine data from the Wide-field Infrared Survey Explorer and Herschel archives to search for Hot DOGs at $z < 0.5$ based on their unique spectral energy distributions. We find 68 candidates, and spectroscopic observations confirm that 3 of them are at $z < 0.5$. For those three, we find their black hole accretion is close to the Eddington limit, with lower bolometric luminosities and black hole masses than those of higher-$z$ Hot DOGs. Compared to high-$z$ systems, these low-$z$ systems are closer to the local relation between host galaxy stellar mass and black hole mass but still lie above it, and we discuss several possible scenarios for it. Finally, we also find the surface number density of $z < 0.5$ Hot DOGs is $2.4 \times 10^{-3}$ deg$^{-2}$, about an order of magnitude lower than high-$z$ Hot DOGs but comparable to hyperluminous unobscured quasars in the same redshift range. These results further support the idea that Hot DOGs may be a transitional phase of galaxy evolution.

## 1. Introduction

In recent decades, a framework of galaxy evolution through gas-rich major mergers has been developed through observations and modeling to explain the coevolution of supermassive black holes (SMBHs) and star formation (e.g., D. B. Sanders et al. 1988; J. E. Barnes & L. Hernquist 1992; F. Schweizer 1998; S. Jogee 2006; A. M. Hopkins & J. F. Beacom 2006; P. F. Hopkins et al. 2008; D. M. Alexander & R. C. Hickox 2012). In this scenario, the tidal torques generated by mergers among gas-rich galaxies funnel large amounts of gas and dust into the center of galaxies, triggering starbursts and rapid, yet highly obscured growth of their SMBHs. The feedback from this active galactic nucleus (AGN) gradually clears the surrounding obscuring material, eventually revealing a visible quasar (e.g., P. F. Hopkins et al. 2008). Throughout the obscured growth phases, these objects may only be bright in the infrared (IR), potentially appearing as submillimeter galaxies (SMGs; A. W. Blain & T. G. Phillips 2002; L. J. Tacconi et al. 2008) and dust-obscured galaxies (DOGs; A. Dey et al. 2008).

Surveying the entire sky in four mid-infrared (MIR) bands centered at 3.4 μm (*W1*), 4.6 μm (*W2*), 12 μm (*W3*), and 22 μm (*W4*), the Wide-field Infrared Survey Explorer (WISE) had among its primary goals to identify the most luminous infrared galaxies (IRGs) in the Universe (E. L. Wright et al. 2010). Using a "W1W2dropout" selection (P. R. M. Eisenhardt et al. 2012), where strong detections are required in *W3* and *W4* but faint or no detection at *W1* and *W2*, a new population of dusty, hyperluminous galaxies was discovered (P. R. M. Eisenhardt et al. 2012; J. Wu et al. 2012). These objects have redshifts peaking at $z \sim 2$–3 (R. J. Assef et al. 2015), with bolometric luminosities ($L_{\rm bol}$) exceeding $10^{13} L_\odot$ (and in ∼10% exceeding $10^{14} L_\odot$; C.-W. Tsai et al. 2015). As they have similar optical-to-mid-IR colors to DOGs but with much higher dust temperatures (⩾60K; J. Wu et al. 2012), these systems are typically referred to as hot dust-obscured galaxies (Hot DOGs).

Follow-up studies revealed that Hot DOGs exhibit a consistent IR spectral energy distribution (SED; J. Wu et al. 2012; C.-W. Tsai et al. 2015), characterized by a broad MIR to far-infrared (FIR) plateau from 10 to 100 $\mu$m in the rest frame, a Rayleigh–Jeans tail at longer wavelengths, and steep drop toward near-infrared (NIR) and optical bands due to high extinction toward the central engine ($A_V \sim$ 20–60, R. J. Assef et al. 2015; and close to Compton thick in the X-rays, D. Stern et al. 2014; E. Piconcelli et al. 2015; R. J. Assef et al. 2016, 2020; C. Ricci et al. 2017; F. Vito et al. 2018). J. Wu et al. (2018) and G. Li et al. (2024) found these systems host SMBHs with masses in the range of $10^8$–$10^{10} M_\odot$, and their Eddington ratios, ranging from 0.1 to 3, are comparable to those of $z \sim 6$ quasars (e.g., R. Wang et al. 2010; G. De Rosa et al. 2011) within the same $M_{BH}$ range. These high Eddington ratios suggest they could be in a transitional "blowout" phase between obscured and unobscured quasars (R. J. Assef et al. 2015; J. Wu et al. 2018; C.-W. Tsai et al. 2018). Furthermore, the recently discovered low-$z$ Hot DOG WISE J190445.04+485308.9 (W1904+4853; G. Li et al. 2023) confirms the existence of Hot DOGs in the local universe, further supporting the hypothetical scenario that they may be a representative population of the "blowout" phase.

While rare, R. J. Assef et al. (2015) found that the number density of Hot DOGs is comparable to that of type-1 quasars with similar luminosities at redshifts $2 < z < 4$. Due to an inherent bias in the selection function, however, there is a significant dearth of Hot DOGs in the $z \leqslant 2$ redshift range. In order to address the absence of Hot DOGs at $z \sim$ 1–2, C. Ricci et al. (2017) devised a complementary WISE color selection function to select Hot DOGs. The extensive follow-up campaign is underway to assess the completeness and contamination of this selection, which will be discussed in R. Assef et al. (2025, in preparation). Naively, one would expect Hot DOGs at $z \lesssim 1$ to be much rarer still given the cosmic evolution of AGN activity, yet their presence at low redshifts remains poorly explored. Identifying Hot DOGs at low redshifts, particularly at $z \lesssim 0.5$, is crucial, as they enable a range of studies that cannot be carried out efficiently at higher-$z$. At this redshift range, the wealth of emission lines covered by optical spectra facilitates the use of BPT diagnostic methods with the advantage of better resolution power on smaller physical scales. Assuming the number density of Hot DOGs is comparable to that of similarly luminous unobscured quasars (R. J. Assef et al. 2015), G. Li et al. (2023) roughly estimated that there are only ∼30 hyperluminous Hot DOGs in the entire sky within the redshift range of 0.36–0.46. In this work, we focus on sources with redshifts less than 0.5 to balance the number of low-$z$ Hot DOGs while retaining as many spectral lines as possible. For convenience, we refer to $z = 0.5$ as the boundary between high-$z$ and low-$z$ Hot DOGs throughout this work.

In this paper, we present a Hot DOGs sample from the Herschel and WISE archive based on their unique SED, focusing on the low-$z$ Hot DOGs. We explore the properties of their host galaxies and central AGNs using optical imaging and spectroscopy, following similar approaches presented in our previous work (G. Li et al. 2023). In Section 2, we describe the sample selection, follow-up spectroscopic observations, and the corresponding multiband photometric data. Section 3 describes the methods for SED fitting, emission-line modeling, and $M_{BH}$ and $M_{star}$ estimations. The results are discussed in Section 4. We summarize our work in Section 5. We adopt a flat ΛCDM cosmology with $H_0 = 70\,\mathrm{km\,s^{-1}\,Mpc^{-1}}$ and $\Omega_m = 0.3$ (G. Hinshaw et al. 2013).

## 2. Sample Selection and Spectral Observation

### 2.1. Sample Selection

We start by considering sources from two FIR photometric databases providing Herschel galaxy photometry: (i) the Herschel-ATLAS survey at the north Galactic pole (∼177 deg$^2$; S. Eales et al. 2010; N. Bourne et al. 2016; E. Valiante et al. 2016), and (ii) the PACS Point Source Catalog (∼1939 deg$^2$; G. Marton et al. 2017) along with the SPIRE Point Source Catalog (∼3383 deg$^2$; B. Schulz et al. 2017). Mirroring the original selection of Hot DOGs (P. R. M. Eisenhardt et al. 2012), we exclude sources within 10° of the Galactic plane or 30° of the Galactic center to avoid contamination from Galactic objects. Additionally, we remove objects from the programs "OT1_peisenha_1" and "OT2_peisenha_2" (PI: P. Eisenhardt), which specifically targeted "W1W2dropout" sources (C.-W. Tsai et al. 2025, in preparation), to enable independent statistical analysis. To constrain their FIR SED, we require candidate sources with at least one marginal detection (⩾3$\sigma$) in the Herschel/PACS bands (70, 100, and 160 $\mu$m; A. Poglitsch et al. 2010) and the Herschel/SPIRE bands (250, 350, and 500 $\mu$m; M. J. Griffin et al. 2010), respectively, resulting in a sample covered by 3% of the sky area (∼1238 deg$^2$). Crossmatching with the WISE all-sky survey catalog (AllWISE; E. L. Wright et al. 2010; R. M. Cutri et al. 2013) and then selecting sources that have marginal detections (3$\sigma$) in all four WISE bands result in an initial sample of 14,526 objects. The requirement for detection in all WISE bands makes this selection generally complementary to the "W1W2dropout" selection of P. R. M. Eisenhardt et al. (2012), although we note they put a magnitude limit on *W1* to favor selection on high-$z$ Hot DOGs. We put a signal-to-noise ratio (SNR) limit instead as depth, which is a strong function of ecliptic latitude in AllWISE that may miss some high-$z$ Hot DOGs.

To constrain the NIR-MIR slope of SED (see Figure 1), we further require that $S_\nu(W1) < S_\nu(W2) < S_\nu(W3)$, where $S_\nu \equiv \nu F_\nu$. We also estimate the optical flux from the UV to 1 $\mu$m ($F_{optical}$) and the total flux ($F_{total}$) for each object by integrating over the observed UV-FIR SED (see Section 2.3 for the additional photometry used). After removing sources with $F_{optical}/F_{total} > 10\%$, as Hot DOGs are dominated by IR emission (P. R. M. Eisenhardt et al. 2012; R. J. Assef et al. 2015; C.-W. Tsai et al. 2015, 2025, in preparation), we visually inspected their optical images and excluded 21 objects that are local mergers or HII regions, remaining as a sample of 366 candidates. Following C.-W. Tsai et al. (2015), we update the scatter of SEDs for "W1W2dropout"-selected Hot DOGs using spectral data from P. R. M. Eisenhardt et al. (2025, in preparation), and horizontally shift and vertically scale the SED of each candidate to confirm if it falls into the SED range, as shown in Figure 1. Given the wavelength limitation of the SED range of high-$z$ Hot DOGs (<160 $\mu$m), we require that at least the UV-Herschel/PACS (70, 100, 160 $\mu$m) SED of each candidate falls into this range. There are 139 objects selected, including 6 manually added objects whose SEDs might be affected by strong silicate absorption and do not fully fall into the SED range. And then we visually selected 68 Hot DOG candidates among them based on their IR SED characteristics, specifically a broader plateau spanning 10–100 $\mu$m, rather than a graybody emission profile. The remaining sources were classified as type 2 AGN and IRGs based on their SED shapes.

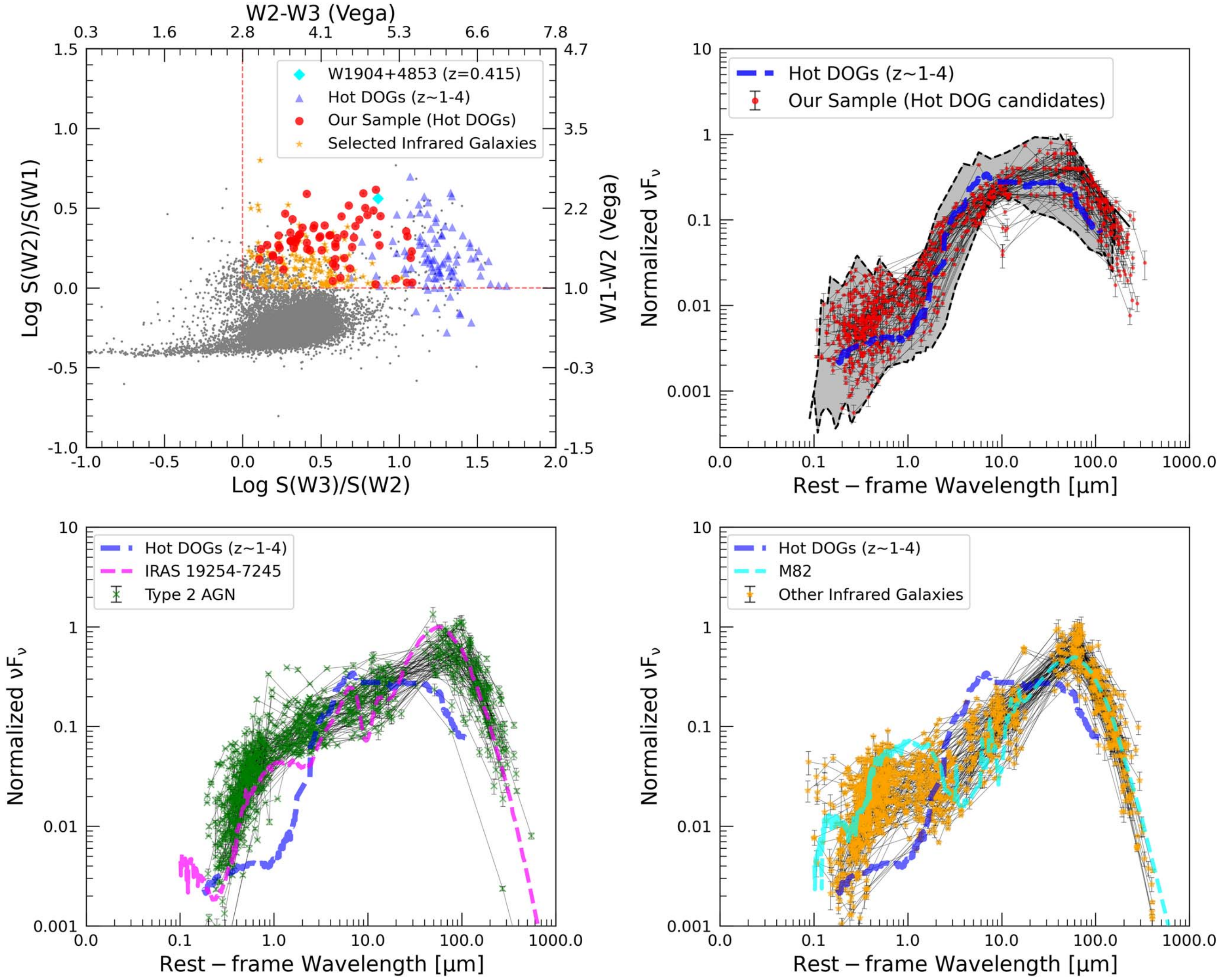

**Figure 1.** Upper left: $S_\nu(W3)/S_\nu(W2)$ vs. $S_\nu(W2)/S_\nu(W1)$. The gray points are the full sample of 14,526. The red points represent the Hot DOG candidates in our sample, while orange stars represent infrared galaxies. The plotted comparative data include W1904+4853 (G. Li et al. 2023) and "W1W2dropout" selected high-$z$ Hot DOGs with Herschel/PACS and Herschel/SPIERE detections (G. Li et al. 2024; C.-W. Tsai et al. 2025, in preparation). Upper right: the SEDs of the Hot DOG candidates. The blue line represents the floating median of SEDs for high-$z$ Hot DOGs (C.-W. Tsai et al. 2025, in preparation). The shaded region in gray represents the updated scatter of SEDs for all "W1W2dropout"-selected Hot DOGs with spectroscopic redshift (P. R. M. Eisenhardt et al. 2025, in preparation). The SED of each candidate is constrained in the shaded area and shifted to the position closest to the median SED. Lower: normalized rest-frame SEDs of objects classified as type 2 AGN (green square) and infrared galaxy (orange star) by our visual inspection. The magenta dashed line represents the SED of IRAS 19254-7245 (M. Polletta et al. 2007), a typical template for a type 2 AGN with a starburst. The cyan dashed line is the SED of the starburst galaxy M82, also from M. Polletta et al. (2007). The SED of individual sources are normalized by the total flux and are shifted to the rest frame based on the photometric redshift as described in Section 2.3.

**Table 1**
Source Selection

| Selection Criteria | Number |
|---|---|
| 1. Initial Sample | 14,526 |
| 2. $S_\nu(W1) < S_\nu(W2)$ and $S_\nu(W2) < S_\nu(W3)$ | 620 |
| 3. $F_{\mathrm{optical}}/F_{\mathrm{total}} < 10\%$ | 387 |
| 4. Visual Selection | 68 (Hot DOG candidates)<br>151 (Infrared galaxies)<br>147 (Type 2 AGNs)<br>21 (Local mergers and HII regions) |

We also visually inspect candidates that do not fall within the SED range. Finally, we classified these 366 candidates into 68 Hot DOGs candidates, 147 type 2 AGNs, and 151 IRGs. The SEDs of these three categories are presented in Figure 1. Our sample selection is detailed in Table 1. For the SED of a selected type 2 AGN, the UV-to-NIR SED indicates it is a type 2 AGN, while the FIR bump likely comes from star formation, similar to IRAS 19254-7245 (see Figure 1), which is thought of as a typical template for a type 2 AGN with a starburst component (M. Polletta et al. 2007). For the selected IRGs, their SEDs are similar to a starburst galaxy template (M82; M. Polletta et al. 2007).

Given the selection criterion (especially at mid-IR bands) is not fully identical to the "W1W2dropout" selection at high-$z$,

**Table 2**
Palomar/DBSP Observation

| ID | Short Name | R.A. (J2000) | Decl. (J2000) | $r$ (mag) | $z_{\rm spec}$ | $z_{\rm phot}$ | Sample | UT |
|---|---|---|---|---|---|---|---|---|
| WISE J005057.25+432038.3 | W0050+4320 | 12.7385461 | 43.3439942 | 18.75 | 0.187 | ... | IRG | 2022/01/03 |
| WISE J094524.16−084007.6 | W0945−0840 | 146.3506869 | −8.6687953 | 19.91 | 0.692 | 0.516 | HD cand | 2022/01/03 |
| WISE J110151.93+571601.8 | W1101+5716 | 165.4664116 | 57.2671921 | 19.53 | 0.354 | 0.401 | IRG | 2022/01/03 |
| WISE J121924.14+164158.8 | W1219+1641 | 184.8505863 | 16.6996928 | 21.13 | 1.105 | 0.633 | HD cand | 2022/01/03 |
| WISE J124915.41+324357.2 | W1249+3243 | 192.3142251 | 32.732575 | 20.38 | 0.679 | 0.404 | HD cand | 2024/04/03 |
| WISE J125327.50+254747.6 | W1253+2547 | 193.3645964 | 25.7965616 | 19.36 | 0.485 | 0.422 | HD cand | 2024/07/05 |
| WISE J132609.90−113107.3 | W1326−1131 | 201.5412802 | −11.5187002 | 21.38 | 1.335 | ... | HD cand | 2022/01/03 |
| WISE J133332.09+503519.6 | W1333+5035 | 203.3837168 | 50.5887918 | 19.93 | 0.801 | 0.635 | HD cand | 2022/01/03 |
| WISE J134617.74+300107.2 | W1346+3001 | 206.5739542 | 30.0186705 | 20.54 | 0.410 | 0.492 | IRG | 2024/04/03 |
| WISE J164250.09+412318.1 | W1642+4123 | 250.7087420 | 41.3883871 | 21.92 | 1.276 | 1.002 | HD cand | 2022/01/03 |
| WISE J170535.38+603456.0 | W1705+6034 | 256.3974538 | 60.5822279 | 21.63 | 1.708 | 0.976 | HD cand | 2022/01/03 |
| WISE J200511.41+021517.6 | W2005+0215 | 301.2975498 | 2.2548958 | 20.81 | 0.782 | ... | HD cand | 2022/01/03 |
| WISE J221159.98−114535.4 | W2211−1145 | 332.9999216 | −11.7598478 | 18.04 | 0.362 | 0.240 | IRG | 2022/01/03 |

**Note.** "HD cand" indicates Hot DOG candidates. "SF" indicates the star-forming galaxies. "IRG" indicates infrared galaxy.

we notice that the SED slope of low-$z$ candidates at mid-IR is in general slightly shallower than the Hot DOG template in Figure 1. This is likely due to the fact that low-$z$ candidates, compared to their high-$z$ counterparts, are slightly less massive and luminous and experience less extinction, which can be attributed to the lower abundance of gas and dust in the local universe. They may also trace less extreme cases of accretion and feedback activities than high-$z$ Hot DOGs that reflect the cosmic evolution history of galaxies. But as long as they represent a similar evolutionary phase as Hot DOGs, with a transition from a hyperluminous, heavily observed, fast accreting AGN to an optical AGN, revealed by their SEDs and optical spectra, we still identified them as good analogs of high-$z$ Hot DOGs at low redshifts. In fact, as shown in Figure 1, the SED of these low-$z$ candidates also falls within the scatter of the high-$z$ Hot DOG SEDs.

### 2.2. Spectroscopic Data

We obtained spectroscopic redshifts ($z_{\rm spec}$) from the Dark Energy Spectroscopic Instrument (DESI; DESI Collaboration et al. 2024a, 2024b), the Sloan Digital Sky Survey (SDSS; Abdurro'uf et al. 2022), and other published data in J. D. Silverman et al. (2010), E. Kalfountzou et al. (2011), A. Hernán--Caballero & E. Hatziminaoglou (2011), I. K. Baldry et al. (2018), and P. R. M. Eisenhardt et al. (2025, in preparation). We also conducted the spectroscopic follow-up observations with Palomar 200 inch Hale Telescope discussed below. The candidates with lower photometric redshifts (see Section 2.3) but no existing spectroscopic redshifts have higher priority in our observations. Currently, 43% (29/68) of candidates have spectroscopic redshifts.

Our spectroscopic follow-up observations were conducted using the Double Spectrograph (DBSP) on the Palomar 200 inch Hale Telescope. These observations were carried out on UT 2021 July 4, 2022 January 3, and 2023 July 1 (PI: Guodong Li), 2024 April 3 (PI: Kevin McCarthy), and 2024 July 5 (PI: Peter R. M. Eisenhardt). We limited the spectra from the blue and red arms of the spectrograph to observed-frame wavelengths 3500–5500Å and 5700–10400 Å respectively, to minimize sensitivity losses at the edges of the observed spectral ranges. Due to weather and other issues, we only obtained spectra of nine candidates from our sample. One of these, W1253+2547 at $z$=0.4836, has spectroscopic observations from SDSS (DR13; F. D. Albareti et al. 2017), but with low SNR at the redder wavelengths. We reobserved this object to detect the H$\alpha$/[N II] lines to be able to better constrain the position of the target in the BPT diagram.

We also observed four objects cataloged as IRGs by our visual inspection. These galaxies are listed in Table 2, and their spectra are shown in Appendix B2. Each spectrum was processed and analyzed with IRAF, following standard reduction procedures to account for bias and flat-fielding. The flux calibration for the spectra was corrected using the standard stars Feige 110, HZ 44, and Wolf 1346.

### 2.3. Multiwavelength Broadband Photometry

We assembled multiwavelength broadband photometry spanning UV to FIR wavelengths to construct the SED of each system. The UV photometry data were obtained from the Galaxy Evolution Explorer (GALEX) source catalogs (M. Seibert et al. 2012). The optical photometry comes from the Legacy Survey (DR10; A. Dey et al. 2019), which includes observations in four bands ($g$, $r$, $i$, and $z$) covering over 15,000 deg$^2$. We also crossmatched our sample with the Pan-STARRS1 survey (PS1; K. C. Chambers et al. 2016; H. A. Flewelling et al. 2020), which covers the 3$\pi$ sr at decl. $> -30°$ in the optical *grizy* bands. For the NIR photometry, we use data from the Two Micron All Sky Survey (2MASS; M. F. Skrutskie et al. 2006) and the UKIRT Infrared Deep Sky Survey (A. Lawrence et al. 2007).

We use the photometric redshifts ($z_{\rm phot}$) from the DESI Legacy Imaging Surveys photometric redshift catalog (A. Dey et al. 2019; R. Zhou et al. 2021), which is developed for luminous red galaxy (LRGs) samples and provides $z_{\rm phot}$ with a normalized median absolute deviation ($\sigma_{\rm NMAD}$) of $\delta z \sim 0.02$ for objects with $z$-band <21 mag. The DESI Legacy Imaging Surveys cover 94% (64) of the candidates in our sample. These photometric redshifts are estimated by the random forest regression method (L. Breiman 2001), a machine learning algorithm based on decision trees. However, since their dust temperature is higher than that of LRGs, Hot DOGs have higher redshifts compared to LRGs with similar IR colors. This suggests that the $z_{\rm phot}$ of Hot DOGs in our sample may be underestimated. There are only seven sources with $z_{\rm phot} < 0.5$ in our sample (see Figure 2). The follow-up identification using spectroscopic data shows that only three of these objects have $z_{spec} < 0.5$.

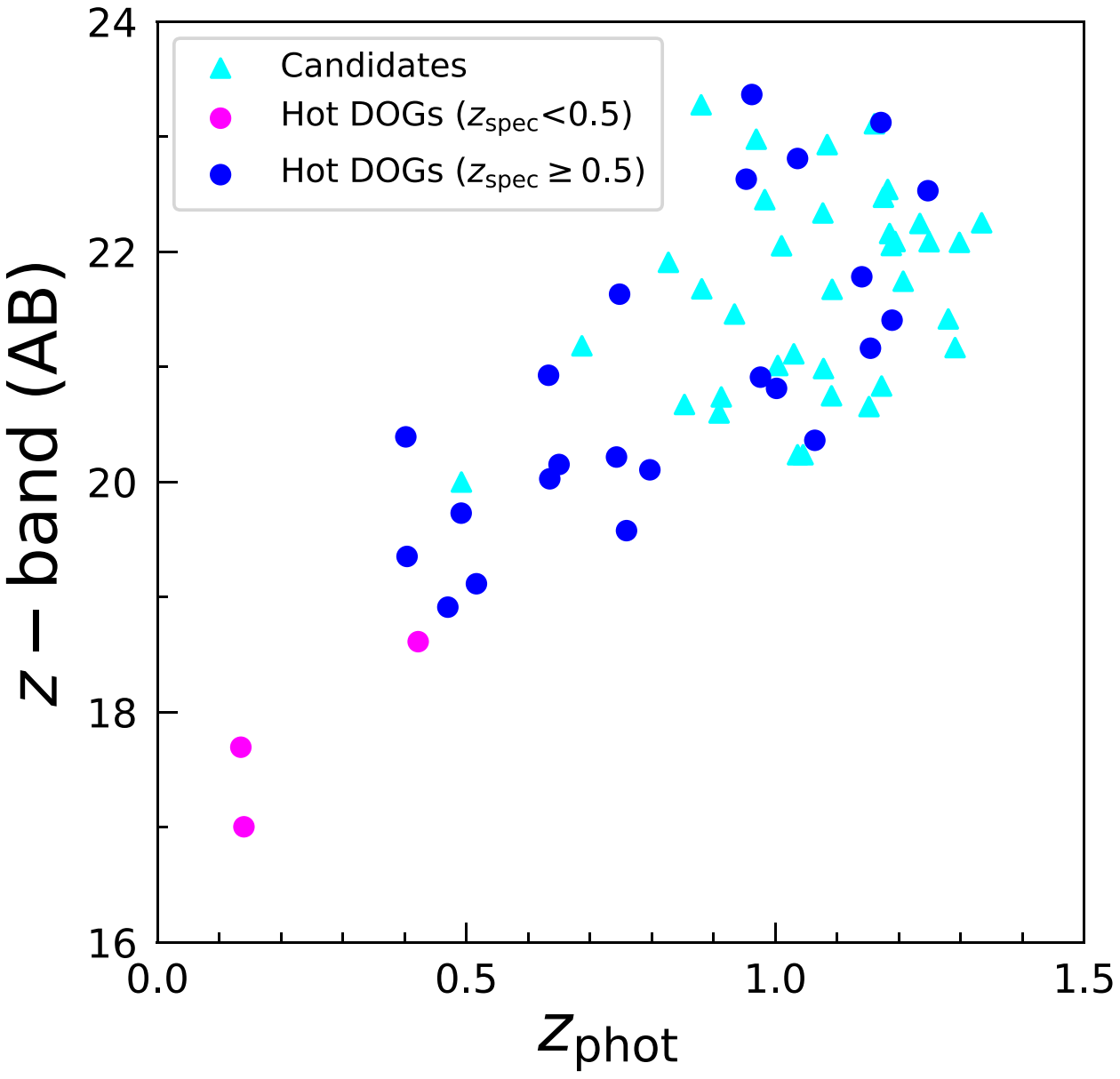

**Figure 2.** $z_{phot}$ vs. $z$-band magnitude for Hot DOG candidates in our sample. The magenta points represent confirmed low-$z$ ($z < 0.5$) Hot DOGs, while blue points are confirmed high-$z$ ($z > 0.5$) Hot DOGs. The cyan points represent the remaining candidates without spectroscopic redshifts. See Section 2.3 for more details.

As discussed by J. Wu et al. (2012) and C.-W. Tsai et al. (2015), high-$z$ Hot DOGs selected by "W1W2dropout" exhibit a consistent IR SED, and their $L_{bol}$ exceeds $10^{13}L_{\odot}$. We identify Hot DOGs if they have a similar SED shape as those extracted from high-$z$ Hot DOGs. Finally, we identified 29 Hot DOGs with spectroscopic redshifts in our sample, including three low-$z$ ($z < 0.5$) systems WISE J091345.49+405628.0, WISE J140638.19+010254.6, and WISE J125327.50+254747.6 (hereafter W0913+4056, W1406+0102, and W1253+2547); see Figure 3 and Figure 4. We note that the low-$z$ Hot DOG W1904+4853 (G. Li et al. 2023) is not included in our sample, as it was not covered by the Herschel survey.

## 3. Method

### 3.1. Photometry Correction

The optical spectra of Hot DOGs often exhibit strong and broad emission lines, along with a faint continuum (J. Wu et al. 2012; G. Li et al. 2023; P. R. M. Eisenhardt et al. 2025, in preparation). In the only $z < 0.5$ Hot DOG studied so far, W1904+4853 (see Figure 1 in G. Li et al. 2023), strong emission lines (Hα, Hβ, [O III], [O II], etc.) dominate the observed broadband fluxes in some filters. When the optical SED is fit using templates with regular equivalent widths for these emission lines, photometric points need to be corrected to remove the excess line contributions. We here apply corrections to the photometry to account for this excess emission, as well as for slit losses, following the same procedure outlined in G. Li et al. (2023). Table 3 lists the multiwavelength luminosity data and line-corrected photometry for the three low-$z$ Hot DOGs ($z < 0.5$) identified in this article, and is shown in Figure 4.

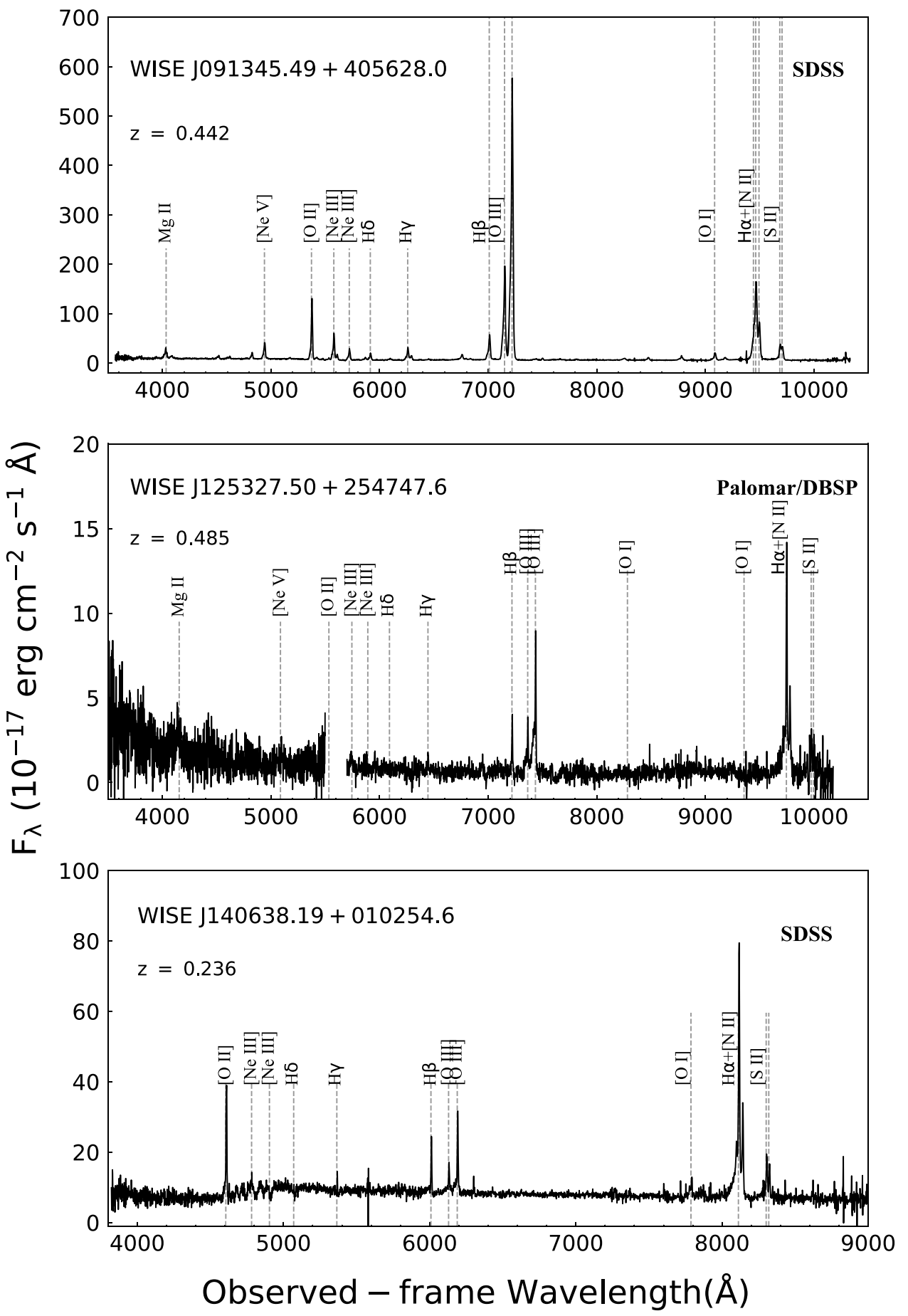

**Figure 3.** The optical spectra of three low-$z$ Hot DOGs.

### 3.2. SED Fitting

The SEDs of high-$z$ Hot DOGs are dominated by a highly obscured AGN (R. J. Assef et al. 2015; C.-W. Tsai et al. 2015), which contributes over 90% to the bolometric luminosity, with nearly all of its emission in the NIR-FIR wavelength (L. Fan et al. 2016; L. Fan et al. 2017, 2018). The recent studies have shown that their FIR luminosity (8–1000 $\mu$m) is AGN dominated, with star formation contributing less than 30%–40% (e.g., L. Finnerty et al. 2020; W. Sun et al. 2024). For the recently discovered low-$z$ Hot DOGs W1904+4853, its star formation contributes only ∼8% to the FIR luminosity (G. Li et al. 2023). To maintain consistency and enable comparison with previous Hot DOGs analyses, we follow the previous method here on the SED analysis, using a minimalistic SED modeling approach to avoid potential parameter degeneracies that can occur when using more detailed approaches (e.g., M. Boquien et al. 2019).

The full UV-FIR SED is decomposed into two parts for concurrent fitting as shown in Figure 4. For the rest UV to NIR SED (<5 $\mu$m), we adopt the SED models and algorithm of R. J. Assef et al. (2010), using a nonnegative linear combination of three empirical galaxy templates (an old stellar population "E" galaxy, an intermediate star-forming "Sbc" galaxy, and a starburst "Im" galaxy). In addition, R. J. Assef et al. (2016) found that a small fraction of Hot DOGs showed UV-optical SEDs too blue to be well explained by any combinations of the galaxy templates but instead required a slightly obscured AGN component, giving rise to the blue Hot

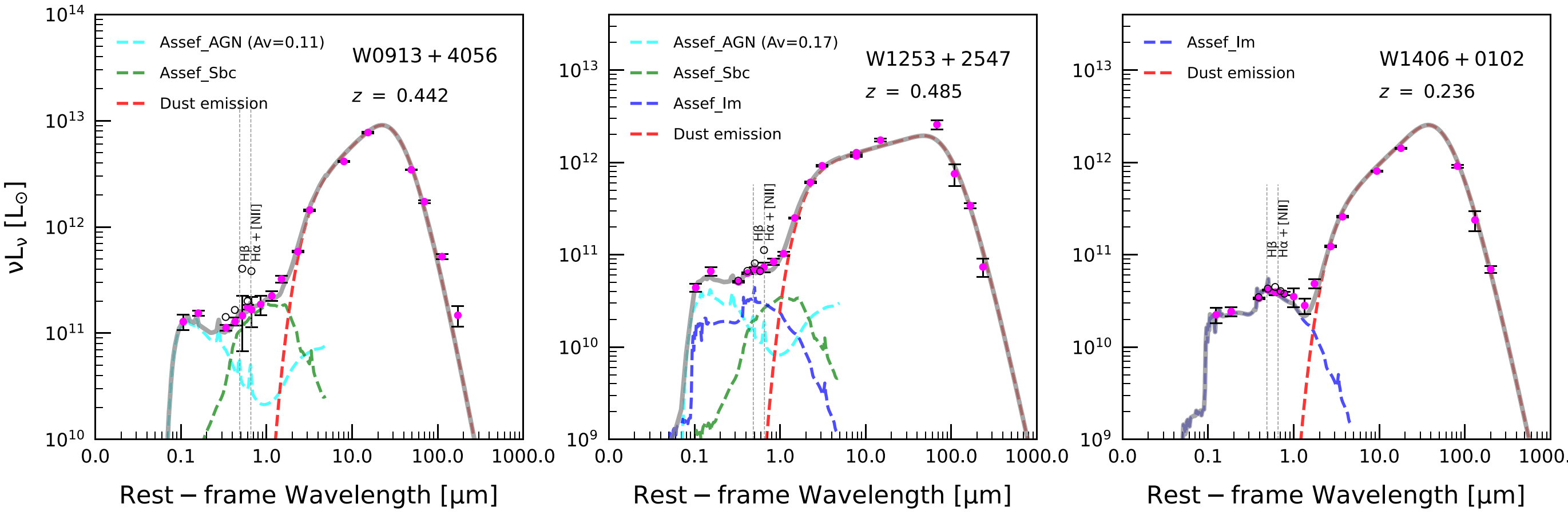

**Figure 4.** SEDs of the three low-$z$ Hot DOGs. The photometric data points are extinction corrected and emission-line subtracted. The black open circles represent the photometry before the emission-line corrections. See Section 3.1 for details. The SED modeling uses templates of empirical galaxies from R. J. Assef et al. (2010), and the single power-law dust distribution model (SPL; red) from C.-W. Tsai et al. (2025, in preparation). The best-fit SED model (heavy gray line) is modeled by a luminous, obscured AGN (dashed cyan line), an intermediate stellar population (dashed blue line), a young stellar population (dashed green line), and the infrared dust model (dashed red line).

**Table 3**
Optical Flux Density of Low-$z$ Hot DOGs

| ID | Band | Flux Density ($\mu$Jy) | Corr Flux ($\mu$Jy) |
|---|---|---|---|
| W0913+4056 | PS1.$g$ | 115 ± 2 | 92 ± 7 |
| | PS1.$r$ | 174 ± 4 | 140 ± 20 |
| | PS1.$i$ | 526 ± 3 | 190 ± 90 |
| | PS1.$z$ | 303 ± 4 | 270 ± 20 |
| | PS1.$y$ | 650 ± 10 | 280 ± 150 |
| W1253+2547 | PS1.$g$ | 35.3 ± 0.8 | 34.3 ± 0.8 |
| | PS1.$r$ | 57.9 ± 0.8 | 55 ± 1 |
| | PS1.$i$ | 86 ± 1 | 74 ± 4 |
| | PS1.$z$ | 82 ± 3 | 82 ± 3 |
| | PS1.$y$ | 150 ± 10 | 100 ± 12 |
| W1406+0102 | PS1.$g$ | 114 ± 3 | 111 ± 3 |
| | PS1.$r$ | 181 ± 2 | 181 ± 3 |
| | PS1.$i$ | 242 ± 3 | 210 ± 20 |
| | PS1.$z$ | 256 ± 5 | 244 ± 9 |
| | PS1.$y$ | 270 ± 10 | 270 ± 10 |

**Note.** "Flux Density" represents the observed value from the Pan-STARRS1 Survey (K. C. Chambers et al. 2016; H. A. Flewelling et al. 2020). "Corr Flux" is the corrected flux, removing the contribution of emission lines using the observed spectra.

DOG (BHD) classification. Thus, we add an AGN component to model the blue excess, as in R. J. Assef et al. (2016, 2020) and G. Li et al. (2024). We also fit for the reddening of the AGN component, assuming $R_V = 3.1$ (see R. J. Assef et al. 2010 for further details). Thus, there are five parameters in the UV-NIR SED modeling, namely, the amplitude of the three host galaxy templates and the amplitude as well as a reddening factor of the AGN component. We initially fit all objects using these four components and then determined the best combination based on the improvement in $\chi^2$ from adding a template that has a $\leqslant$10% chance of being spurious according to an F-test.

The IR SED ($\lambda_{\rm rest} > 5\ \mu$m) is fitted by a single power-law (SPL) dust distribution model from C.-W. Tsai et al. (2025, in preparation), which was developed to model the IR SEDs of high-$z$ Hot DOGs and can overcome the limitations of empirical templates from R. J. Assef et al. (2010), which only extend to 30 $\mu$m. The IR SEDs of Hot DOGs cannot be well modeled by the torus model, as they rise more sharply at mid-IR wavelengths and fall into the Rayleigh–Jeans regime at a shorter wavelength than the Torus model (see C.-W. Tsai et al. 2015). Other results using models like CLUMPY (M. Nenkova et al. 2008) suggest that the dust cocoon in Hot DOGs almost spherically enshrouds the ultraluminous source within these galaxies (C.-W. Tsai et al. 2025, in preparation). The SEDs of Hot DOGs do not match any previously known IR-luminous systems. Inspired by their unique SED characteristics, C.-W. Tsai et al. (2025, in preparation) have developed an SPL model to describe their dust emission. The SPL model assumes a continuous set of spherically symmetric, optically thin dust shells in thermal equilibrium surrounding the central AGN, where the density of each layer is a power-law function of the radius from the galaxy center. Assuming a dust emissivity $\beta = 1.5$, the parameters for this model are the scaling factor, the power-law index $p$ of the dust density, the representative hottest dust temperature ($T_{\rm hot}$) of the inner cutoff of the dust distribution, and the representative coldest dust temperature ($T_{\rm cold}$) of the outer boundary of the dust distribution.

The optical-NIR and IR SEDs are fitted simultaneously using a Markov Chain Monte Carlo (MCMC) method to more efficiently sample the parameter space. Specifically, we use the MCMC implementation of D. Foreman-Mackey et al. (2013) through the public Python package `emcee`.[18] We use uniform priors for all model parameters, enforcing all to be nonnegative. The median of the marginal distribution of each parameter is taken as its best-fit value, and the quoted 1$\sigma$ uncertainties correspond to the 16th and 84th percentiles of these distributions. Figure 4 shows the best-fit SED models of these three low-$z$ Hot DOGs. We also performed a similar SED fitting on high-$z$ Hot DOGs in our sample, without making photometric corrections because most of the strong emission lines mentioned in Section 3.1 have shifted out of the optical wavelengths. We only consider the UV to NIR wavelengths when fitting the objects classified as IRGs by our visual inspection.

[18] https://emcee.readthedocs.io/

### 3.3. Emission-line Fitting

We used a power law to model the continuum emission, and single or double Gaussian profiles to model the emission lines. For lines with significantly blueshifted and asymmetrical profiles, such as forbidden lines or Balmer lines in Hot DOGs (J. Wu et al. 2018; H. D. Jun et al. 2020; G. Li et al. 2023), one Gaussian is used to fit the broad and blueshifted component, while another is used to model the narrow component. For lines with symmetric structures, a second Gaussian was only added when an F-test showed a low probability (<10%) of the improvement in $\chi^2$ being spurious. Considering that the Mg II line in Hot DOGs is less affected by the Fe II emission complex (C.-W. Tsai et al. 2018; G. Li et al. 2023), we fit it using a double/single Gaussian model without using an Fe II template. For the low-$z$ Hot DOGs with spectroscopic data, we could not measure the broad [N II] emission due to its blending with the H$\alpha$ line, as shown in Figure 3. Thus, we do not include the broad [N II] in our analysis. Table 4 summarizes the emission-line fitting results for the three low-$z$ Hot DOGs, with fitting results for other galaxies observed by Palomar/DBSP provided in Appendix B1.

### 3.4. Luminosity Estimates

Following past works for Hot DOGs (C.-W. Tsai et al. 2015, 2018; J. Wu et al. 2018; G. Li et al. 2023), we estimate the $L_{\rm bol}$ by integrating over the observed UV-FIR SEDs using power-law interpolation between photometric data points, and neglecting the contribution from shorter or longer wavelengths. We note that both the AGN and host galaxy contributions are taken into account in the estimation of the bolometric luminosity, with the AGN contributing over 98%. The systematic uncertainty of the total luminosity is estimated to be $\sim$20% based on the assumption of a slow-varying SED (C.-W. Tsai et al. 2015).

### 3.5. Stellar Mass Estimates

We estimate $M_{\rm star}$ using the correlations between the rest-frame $g-r$ color and the rest-frame $H$-band $M/L_{\rm H}$ ratio reported by E. F. Bell et al. (2003). The $M/L_{\rm H}$ ratio is given by $\log_{10}(M/L_H) = -0.189 + (0.266 \times (g-r))$. To estimate the host galaxy fluxes in these three bands, we use the best-fit SED model and remove the AGN component and dust emission from SPL model as described in Section 3.2. We note that, while the host-galaxy $g-r$ color is poorly constrained in cases where AGN scattered light may be present, the $H$-band mass-to-light ration is not very sensitive to this color, varying by a factor of up to 2 across diverse star formation histories (see E. F. Bell et al. 2003). Stellar masses for the three low-$z$ Hot DOGs are listed in Table 5, while the results for high-$z$ Hot DOGs and other IRGs are provided in Appendices A1 and A2.

We note that here we assume no significant dust obscuration in the host galaxies, as discussed in R. J. Assef et al. (2015). High-resolution Atacama Large Millimeter/submillimeter Array (ALMA) images of Hot DOGs ([C II]) reveal that most dust emission comes from the nucleus with minimal contribution from larger scales (T. Dìaz-Santos et al. 2021), while corresponding UV images in R. J. Assef et al. (2020) show a similar extent to the [C II] emission, indicating that large-scale emission might not be significantly obscured. Furthermore, a recent spatial dust emission analysis of the most luminous Hot DOG, W2246-0526, using ALMA indicates that its nuclear region is optically thick, while larger-scale regions are optically thin (R. Fernández Aranda et al. 2025), further supporting our assumption.

### 3.6. Black Hole Mass Estimates

Assuming that the broad emission lines are from the virialized gas around the central SMBH (e.g., A. Wandel et al. 1999; C. A. Onken et al. 2004), their broad-line widths trace the velocity dispersion of the gas and can be utilized to estimate the black hole (BH) mass ($M_{\rm BH}$). This method relies on the empirical relationship between the broad-line region size and AGN luminosity ($R$–$L$ relation) calibrated through reverberation mapping studies, typically for the H$\beta$ emission line and the continuum luminosity at 5100 Å (e.g., S. Kaspi et al. 2000, 2005; B. M. Peterson et al. 2004; M. C. Bentz et al. 2009, 2013). The broad H$\alpha$ and Mg II line width and luminosity correlate well with those of H$\beta$ over a wide range of total AGN luminosities (e.g., R. J. McLure & M. J. Jarvis 2002; J. E. Greene & L. C. Ho 2005; M. Vestergaard & P. S. Osmer 2009; J.-G. Wang et al. 2009; B. Trakhtenbrot et al. 2011; H. D. Jun et al. 2015), supporting the use of those lines when available, adjusting the $R$–$L$ relation for those specific emission lines. The C IV line is also commonly used to estimate $M_{\rm BH}$ in high-$z$ AGN, although its line width correlates less well with H$\beta$ or H$\alpha$ and often displays asymmetric, blueshifted profiles that potentially require correction terms (e.g., M. Vestergaard & B. M. Peterson 2006; Y. Shen & X. Liu 2012; D. Park et al. 2013; J. E. Mejìa-Restrepo et al. 2016; L. Coatman et al. 2016, 2017).

Hot DOG's Balmer lines are often broad and blueshifted (J. Wu et al. 2018; H. D. Jun et al. 2020; G. Li et al. 2023), indicating that the emitting gas may be outflowing, and may not always be suitable for $M_{\rm BH}$ estimates. In contrast, the Mg II line observed in the discovered low-$z$ Hot DOGs displays a symmetric broad profile with less blueshift (G. Li et al. 2023), suggesting that Mg II does not appear to be significantly affected by outflows and might originate from gas around the SMBH and escape through scattering off the dust cocoon (R. J. Assef et al. 2016, 2020, 2022). Moreover, through statistical analysis of high-$z$ Hot DOGs, G. Li et al. (2024) also found that the broad Mg II and C IV lines in Hot DOGs might come from the scattered light of the central AGN. Thus, following G. Li et al. (2023), for low-$z$ Hot DOGs, we estimate the SMBH mass using the Mg II width and the formula from Equation (10) of J.-G. Wang et al. (2009):

$$\log\left(\frac{M_{\rm BH}}{M_\odot}\right) = (7.13 \pm 0.27) + 0.5\log\left(\frac{L_{3000}}{10^{44}\ \ {\rm erg\ s^{-1}}}\right) + (1.51 \pm 0.49)\log\left(\frac{{\rm FWHM_{MgII}}}{1000\ {\rm km\ s^{-1}}}\right), \quad (1)$$

where $L_{3000}$ is the monochromatic luminosity at rest-frame 3000 Å. The UV luminosity of Hot DOGs cannot be directly measured because of the high internal UV/optical extinction toward its central region. Therefore, we estimate $L_{3000}$ using $L_{3000} \sim 0.19 \times L_{\rm bol}$ based on the unobscured AGN template of G. T. Richards et al. (2006). We note that the Mg II line for W1406+0102 is not covered by the optical spectrum due to its lower redshift. Considering the lower blueshift of the broad H$\alpha$ line ($-210$ km s$^{-1}$) compared to H$\beta$ as shown in Table 4, we estimate $M_{\rm BH}$ using the broad H$\alpha$ line, following

**Table 4**
Emission-line Properties of Low-$z$ Hot DOGs

| Name | Line | $\log L_{\rm line}$ (erg s$^{-1}$) | $\log L_{\rm bl}$ (erg s$^{-1}$) | $\Delta v_{bl}$ (km s$^{-1}$) | FWHM$_{\rm bl}$ (km s$^{-1}$) | EW$_{\rm bl}$ (Å) | $\log L_{\rm nl}$ (erg s$^{-1}$) | $\Delta v_{\rm nl}$ (km s$^{-1}$) | FWHM$_{\rm nl}$ (km s$^{-1}$) | EW$_{\rm nl}$ (Å) |
|---|---|---|---|---|---|---|---|---|---|---|
| W0913+4056 | Mg II | 42.56 ± 0.02 | 42.51 ± 0.01 | −130 ± 20 | 2310 ± 50 | 32 ± 1 | 41.60 ± 0.10 | −80 ± 10 | 380 ± 30 | 4 ± 1 |
| | [Ne V]λ3346 | 42.16 ± 0.02 | 41.64 ± 0.05 | −690 ± 70 | 850 ± 90 | 4 ± 1 | 42.01 ± 0.02 | 30 ± 20 | 630 ± 30 | 10 ± 1 |
| | [Ne V]λ3426 | 42.60 ± 0.01 | 41.98 ± 0.02 | −750 ± 30 | 840 ± 40 | 9 ± 1 | 42.48 ± 0.01 | 46 ± 7 | 700 ± 10 | 28 ± 1 |
| | [O II] | 43.11 ± 0.01 | 42.34 ± 0.02 | −870 ± 20 | 900 ± 30 | 25 ± 1 | 43.04 ± 0.01 | −86 ± 3 | 634 ± 4 | 124 ± 1 |
| | Hζ | 42.85 ± 0.01 | 42.66 ± 0.01 | −730 ± 30 | 1390 ± 10 | 63 ± 1 | 42.39 ± 0.02 | 20 ± 6 | 461 ± 8 | 34 ± 1 |
| | Hδ | 42.28 ± 0.02 | 41.98 ± 0.02 | −410 ± 50 | 1570 ± 90 | 14 ± 1 | 41.98 ± 0.02 | 18 ± 7 | 550 ± 20 | 15 ± 1 |
| | Hγ | 42.60 ± 0.01 | 42.44 ± 0.01 | −120 ± 10 | 1660 ± 30 | 42 ± 1 | 42.10 ± 0.02 | 28 ± 4 | 480 ± 10 | 20 ± 1 |
| | [O III]λ4363 | 42.01 ± 0.03 | … | … | … | … | 42.01 ± 0.03 | 3 ± 9 | 700 ± 20 | 16 ± 1 |
| | Hβ | 42.90 ± 0.01 | 42.63 ± 0.01 | −337 ± 8 | 1350 ± 10 | 55 ± 1 | 42.58 ± 0.01 | 40 ± 2 | 521 ± 6 | 49 ± 1 |
| | [O III]λ4959 | 43.51 ± 0.01 | 43.32 ± 0.01 | −299 ± 2 | 1317 ± 4 | 263 ± 2 | 43.06 ± 0.01 | 28 ± 2 | 470 ± 3 | 146 ± 2 |
| | [O III]λ5007 | 43.98 ± 0.01 | 43.76 ± 0.01 | −363 ± 2 | 1274 ± 3 | 730 ± 10 | 43.58 ± 0.01 | 38 ± 2 | 514 ± 4 | 481 ± 9 |
| | [O I]λ6300 | 42.42 ± 0.01 | 42.13 ± 0.02 | −170 ± 20 | 1320 ± 80 | 21 ± 1 | 42.12 ± 0.02 | 60 ± 5 | 490 ± 20 | 21 ± 1 |
| | Hα | 42.40 ± 0.03 | 42.21 ± 0.05 | −200 ± 20 | 2400 ± 300 | 170 ± 40 | 41.97 ± 0.02 | 26 ± 4 | 246 ± 9 | 100 ± 10 |
| | [N II]λ6548 | 40.5 ± 0.3 | … | … | … | … | 40.5 ± 0.3 | 35 ± 5 | 260 ± 30 | 3 ± 2 |
| | [N II]λ6584 | 41.50 ± 0.05 | … | … | … | … | 41.50 ± 0.05 | 35 ± 5 | 260 ± 30 | 34 ± 5 |
| | [S II]λ6716 | 41.34 ± 0.08 | … | … | … | … | 41.34 ± 0.08 | 70 ± 10 | 360 ± 70 | 21 ± 4 |
| | [S II]λ6731 | 41.20 ± 0.08 | … | … | … | … | 41.20 ± 0.08 | 73 ± 8 | 270 ± 70 | 15 ± 3 |
| W1253+2547 | Mg II | 41.60 ± 0.08 | 41.60 ± 0.08 | 0 ± 90 | 1700 ± 500 | 14 ± 3 | … | … | … | … |
| | [O II] | 42.28 ± 0.03 | 41.91 ± 0.04 | −350 ± 40 | 1050 ± 70 | 18 ± 2 | 42.03 ± 0.04 | 216 ± 7 | 450 ± 20 | 23 ± 2 |
| | [Ne III]λ3868 | 41.92 ± 0.06 | 41.92 ± 0.06 | −370 ± 50 | 1890 ± 110 | 16 ± 1 | … | … | … | … |
| | Hγ | 41.30 ± 0.10 | … | … | … | … | 41.30 ± 0.10 | 79 ± 9 | 280 ± 20 | 4 ± 1 |
| | Hβ | 41.86 ± 0.03 | 41.68 ± 0.04 | −450 ± 50 | 1240 ± 90 | 10 ± 1 | 41.38 ± 0.05 | 90 ± 20 | 340 ± 60 | 27 ± 5 |
| | [O III]λ4959 | 41.80 ± 0.08 | 41.7 ± 0.1 | −200 ± 200 | 2009 ± 5 | 60 ± 20 | 41.1 ± 0.1 | 60 ± 20 | 220 ± 50 | 16 ± 5 |
| | [O III]λ5007 | 42.11 ± 0.04 | 41.92 ± 0.05 | −550 ± 60 | 1300 ± 100 | 100 ± 20 | 41.65 ± 0.05 | 83 ± 7 | 260 ± 20 | 60 ± 10 |
| | Hα | 42.15 ± 0.02 | 42.20 ± 0.05 | −240 ± 70 | 2400 ± 300 | 170 ± 40 | 41.97 ± 0.01 | 80 ± 3 | 250 ± 10 | 100 ± 15 |
| | [N II]λ6548 | 40.5 ± 0.3 | … | … | … | … | 40.5 ± 0.3 | 70 ± 10 | 270 ± 30 | 4 ± 2 |
| | [N II]λ6584 | 41.51 ± 0.05 | … | … | … | … | 41.51 ± 0.05 | 70 ± 10 | 270 ± 30 | 35 ± 7 |
| | [S II]λ6716 | 41.34 ± 0.08 | … | … | … | … | 41.34 ± 0.08 | 110 ± 30 | 400 ± 100 | 21 ± 5 |
| | [S II]λ6731 | 41.20 ± 0.08 | … | … | … | … | 41.20 ± 0.08 | 120 ± 20 | 270 ± 60 | 15 ± 4 |
| W1406+0102 | [O II] | 41.73 ± 0.03 | 41.00 ± 0.10 | −750 ± 70 | 950 ± 120 | 6 ± 2 | 41.64 ± 0.03 | −104 ± 4 | 430 ± 10 | 25 ± 2 |
| | Hγ | 40.64 ± 0.09 | … | … | … | … | 40.64 ± 0.09 | 0 ± 8 | 180 ± 20 | 2.2 ± 0.4 |
| | Hβ | 41.44 ± 0.04 | 41.06 ± 0.07 | −1200 ± 200 | 2000 ± 350 | 6 ± 1 | 41.22 ± 0.05 | 3 ± 4 | 266 ± 9 | 9 ± 1 |
| | [O III]λ4959 | 41.02 ± 0.08 | … | … | … | … | 41.02 ± 0.08 | 0 ± 7 | 350 ± 20 | 5 ± 1 |
| | [O III]λ5007 | 41.60 ± 0.03 | 41.15 ± 0.06 | −600 ± 60 | 1500 ± 120 | 7 ± 1 | 41.41 ± 0.03 | 3 ± 3 | 273 ± 9 | 4 ± 2 |
| | Hα | 42.31 ± 0.01 | 42.08 ± 0.01 | −370 ± 20 | 2200 ± 90 | 74 ± 2 | 41.93 ± 0.01 | 6 ± 1 | 266 ± 5 | 52 ± 2 |
| | [N II]λ6548 | 40.80 ± 0.20 | … | … | … | … | 40.80 ± 0.20 | 9 ± 3 | 273 ± 9 | 4 ± 2 |
| | [N II]λ6584 | 41.50 ± 0.04 | … | … | … | … | 41.50 ± 0.04 | 9 ± 3 | 271 ± 9 | 19 ± 2 |
| | [S II]λ6716 | 41.37 ± 0.07 | … | … | … | … | 41.37 ± 0.07 | 23 ± 8 | 390 ± 20 | 14 ± 2 |
| | [S II]λ6731 | 41.10 ± 0.10 | … | … | … | … | 41.10 ± 0.10 | 59 ± 6 | 290 ± 20 | 8 ± 2 |

**Note.** $L_{\rm line}$ is the line luminosity (narrow+broad) derived from the best-fit model, and $\Delta v$ indicates the offset of the Gaussian center for each broad/narrow component. Subscripts "bl" and "nl" indicate broad line and narrow line, respectively. FWHM and EW are calculated in the rest frame.

**Table 5**
Physical Properties of Low-$z$ Hot DOGs

| Name | Redshift | $\log L_{\rm bol}$ ($L_\odot$) | $T_{\rm Max}$ (K) | $T_{\rm Min}$ (K) | $\log M_{\rm dust}$ ($M_\odot$) | $\log M_{\rm star}$ ($M_\odot$) | $\log M_{\rm BH}$ ($M_\odot$) | $\log \eta_{\rm Edd}$ |
|---|---|---|---|---|---|---|---|---|
| W0913+4056 | 0.442 | 13.2 ± 0.1 | 850 ± 20 | 72 ± 1 | 6.5 | 11.3 | 8.7 ± 0.3 | 0 ± 0.3 |
| W1253+2547 | 0.485 | 12.8 ± 0.1 | 1340 ± 20 | 29 ± 1 | 7.9 | 10.6 | 8.3 ± 0.4 | 0 ± 0.4 |
| W1406+0102 | 0.236 | 12.6 ± 0.1 | 820 ± 20 | 45 ± 1 | 7.1 | 10.2 | 8.3 ± 0.2 | −0.2 ± 0.2 |

**Note.** Redshifts of W0913+4056 and W1406+0102 are adopted from SDSS (Abdurro'uf et al. 2022), while the redshift of W1253+2547 is derived from the Palomar/DBSP observation. The dust temperature limit of "$T_{\rm Max}$," "$T_{\rm Min}$," and the dust mass ($M_{\rm dust}$) are estimated for the SPL model based on C.-W. Tsai et al. (2025, in preparation). Other physical parameters are calculated as described in Section 3.

R. J. Assef et al. (2011) and J. Wu et al. (2018). The results are listed in Table 5.

For high-$z$ Hot DOGs where the Mg II line either is undetected or has shifted out of the observed wavelength range, we correct the FWHM of C IV and estimate $M_{\rm BH}$ using the methodology of L. Coatman et al. (2017; see C.-W. Tsai et al. 2018 for more details). For the selected IRGs, we preferentially use the FWHM of Hβ line to calculate $M_{\rm BH}$ (J.-G. Wang et al. 2009). We provide these $M_{\rm BH}$ estimates in Appendix A (see Table A1 and A2) for reference; however, caution should be exercised when comparing values derived from different lines, as there may be significant systematic differences between the estimators upon the selection of the line or the line width profile (factors of few, e.g., H. D. Jun et al. 2015; E. Dalla Bontà et al. 2020).

## 4. Discussion

### 4.1. UV-FIR Emission for Low-z Hot DOGs

The IR SED shape of low-$z$ Hot DOGs is similar to that of high-$z$ Hot DOGs presented in C.-W. Tsai et al. (2015), as shown in Figure 4. Their dust emission contributes over 98% of the bolometric luminosity, and their optical spectra exhibit strong and broad emission lines. Using the SPL model from C.-W. Tsai et al. (2025, in preparation) to fit the NIR to FIR SED (see Section 3.2), the dust temperatures and dust mass for low-$z$ systems (Table 5) are similar to high-$z$ Hot DOGs (C.-W. Tsai et al. 2025, in preparation). The majority of the difference is in their lower bolometric luminosities, which may be related to the selection effects between low-$z$ and high-$z$ Hot DOG samples. Less luminous Hot DOGs, with BH accretion rates closing the Eddington limit but smaller $M_{\rm BHs}$, can be identified at low redshifts but are undetectable at higher redshifts with WISE and Herschel. We also cannot rule out the possibility that this difference is related to the decline in gas surface density and merger rates in the local universe (e.g., T. K. Garratt et al. 2021; S. Berta et al. 2013; R. Decarli et al. 2016, 2019, 2020; D. A. Riechers et al. 2019, 2020).

Following the identification method of G. Li et al. (2024), W0913+4056 is classified as a low-$z$ BHD, with 60% of its optical emission contributed by a slightly obscured AGN ($A_{\rm V} = 0.11$). In contrast, W1253+2547 and W1406+0102 are classified as "regular" Hot DOGs, whose optical flux is dominated by their host galaxies. Albeit subject to small number statistics, the BHD fraction in low-$z$ Hot DOGs appears to be 33%, which is consistent with the proportion found in high-$z$ Hot DOGs (26%; G. Li et al. 2024). The optical luminosity of W0913+4056 is higher than that for W1253+2547 and W1406+0102. Furthermore, the optical spectra of W1253+2547 and W1406+0102 show clear balmer breaks, but this is absent in W0913+4056. This aligns with the evolutionary scenario that BHDs represent a stage closer to traditional type 1 AGNs in the evolution of Hot DOGs, as discussed in R. J. Assef et al. (2022) and G. Li et al. (2024). In that model, at this evolutionary stage, the dust or gas enveloping the AGN is partially dispersed, allowing light from the AGN to be scattered rather than being fully absorbed by the thick dust, resulting in higher optical emission in BHDs and making continuum spectrum features of their host galaxies undetected.

W0913+4056 is also one of the brightest objects discovered in the IRAS survey (IRAS 09104+4109; S. G. Kleinmann et al. 1988) and has been extensively studied from radio to high-energy X-rays (D. C. Hines & B. J. Wills 1993; A. C. Fabian & C. S. Crawford 1995; L. Armus et al. 1999; D. C. Hines et al. 1999; H. D. Tran et al. 2000; K. Iwasawa et al. 2001; J. R. Deane & N. Trentham 2001; E. Piconcelli et al. 2007; C. Bildfell et al. 2008; A. Ruiz et al. 2010; C. Vignali et al. 2011; J. Hlavacek-Larrondo et al. 2012; F. Combes et al. 2011; E. O'Sullivan et al. 2012; D. Farrah et al. 2016; E. O'Sullivan et al. 2021). Located at the center of the rich cluster MACS J0913.7+4056 (S. G. Kleinmann et al. 1988; P. B. Hall et al. 1997; D. Farrah et al. 2004), W0913+4056 is accreting surrounding small galaxies (J. B. Hutchings & S. G. Neff 1988). It is similar to the most luminous Hot DOG WISE J224607.56-052634.9, which is also believed to be at the center of a protocluster (T. Dìaz-Santos et al. 2018; D. Zewdie et al. 2023; Y. Luo et al. 2024). The surrounding small galaxies provide rich fuel for accretion onto its central BH, possibly explaining its higher luminosity compared to the other two low-$z$ Hot DOGs. Like high-$z$ Hot DOGs (D. Stern et al. 2014; E. Piconcelli et al. 2015; R. J. Assef et al. 2016, 2020; C. Ricci et al. 2017; F. Vito et al. 2018), W0913+4056 is also Compton thick ($N_{\rm H} = 5 \times 10^{23}\,{\rm cm}^{-2}$; F. A. Harrison et al. 2013; G. B. Lansbury et al. 2014, 2015; D. Farrah et al. 2016). It is classified as a "radio-intermediate" FR-I source, with a linear core and double-lobed structure (D. C. Hines & B. J. Wills 1993; E. O'Sullivan et al. 2012). Although highly obscured, it shows broad Hβ, Hγ, and Mg II lines in polarized light (D. C. Hines & B. J. Wills 1993; H. D. Tran et al. 2000), supporting the conjecture about the origin of emission lines in BHDs (R. J. Assef et al. 2016, 2020, 2022; G. Li et al. 2024).

### 4.2. Eddington Ratio and Black Hole Accretion

Spectral line analyses of high-$z$ Hot DOGs reveal that the $M_{\rm BH}$ of these systems range from $10^8$–$10^{10} M_\odot$, and their Eddington ratios are close to or even exceed the Eddington limit (J. Wu et al. 2018; C.-W. Tsai et al. 2018; L. Finnerty et al. 2020; G. Li et al. 2024), comparable to bright quasars at $z \sim 6$ (e.g., L. Jiang et al. 2007; C. J. Willott et al. 2010;

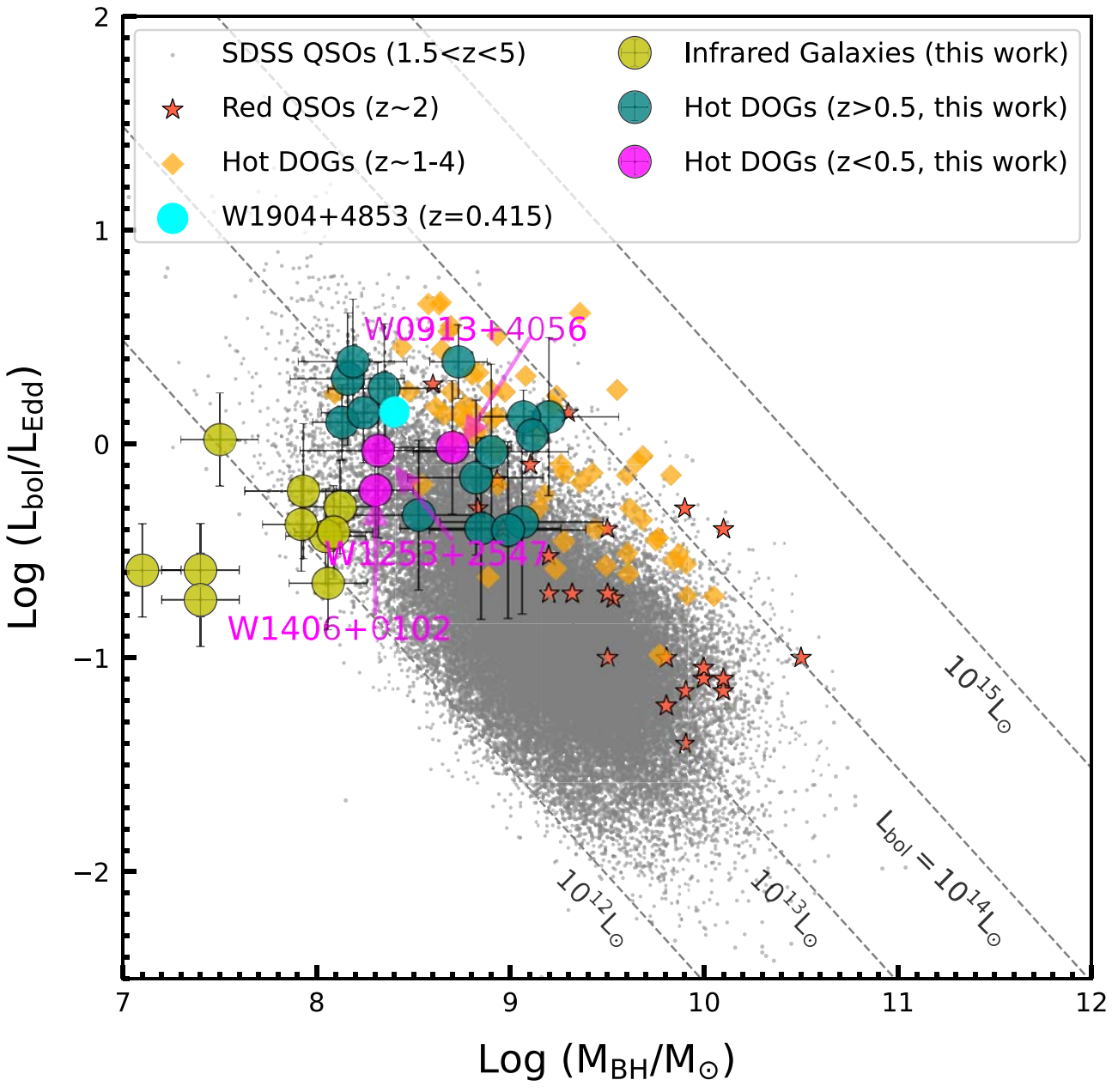

**Figure 5.** Eddington ratio vs. $M_{BH}$ for unobscured and obscured quasars, following G. Li et al. (2024), J. Wu et al. (2018), and C.-W. Tsai et al. (2018). The points with error bars represent the low-$z$ Hot DOGs, high-$z$ Hot DOGs, and low-$z$ IRGs in this work. The plotted comparative data include high-$z$ Hot DOGs (orange stars) selected by "W1W2dropout" (G. Li et al. 2024), SDSS quasars in the range of $1.5 < z < 5$ (Y. Shen et al. 2011), heavily reddened quasars (M. Banerji et al. 2012, 2015), and a Hot DOG at $z \sim 0.415$ (G. Li et al. 2023).

G. De Rosa et al. 2011; C. Mazzucchelli et al. 2023). Figure 5 shows the $M_{BH}$ of low-$z$ Hot DOGs estimated as discussed in Section 3.6. Compared to high-$z$ Hot DOGs, these low-$z$ systems have lower $M_{BH}$ and $L_{bol}$ but with similar Eddington ratios.

We also estimated the $M_{BH}$ of objects classified as IRGs using the same approach, whose optical spectra show broad emission lines. Although these IRGs have lower $M_{BH}$ than that of low-$z$ Hot DOGs, their Eddington ratios (∼0.2–0.6) suggest that these sources are like AGN-dominated systems, such as DOGs (A. Dey et al. 2008). This is consistent with the scenario of Hot DOGs described above (J. Wu et al. 2012; R. J. Assef et al. 2015; J. Wu et al. 2018; C.-W. Tsai et al. 2018), wherein DOGs transition to the Hot DOG phase with increasing $M_{BH}$ and accretion rate, eventually evolving toward traditional type 1 AGNs, as described by P. F. Hopkins et al. (2008).

### *4.3. Emission-line Diagnosis*

The BPT diagram serves as a method for identifying emission-line ionization sources based on line ratios (J. A. Baldwin et al. 1981; L. J. Kewley et al. 2006). As shown in Figure 6, we employ the ratios of [N II]/Hα and [O III]/Hβ (L. J. Kewley et al. 2001, 2006; G. Kauffmann et al. 2003) to distinguish the dominating ionizing processes of low-$z$ Hot DOGs within this work. Located within the AGN region on the BPT diagrams, W0913+4056 is identified as a typical AGN. This classification is consistent with the SED fitting, indicating that the AGN dominates the optical emission. W1253+2547 and W1406+0102 are located, instead, in the composite region between AGN and star-forming galaxies on the BPT diagram, similar to the recently discovered low-$z$ Hot DOG W1904+4853 (G. Li et al. 2023), which also has an optical SED dominated by a star-forming host galaxy. This suggests that such regular Hot DOGs may harbor strong starburst activity (also see discussion in S. F. Jones et al. 2014; and W. Sun et al. 2024), placing them in the composite region of the BPT diagram.

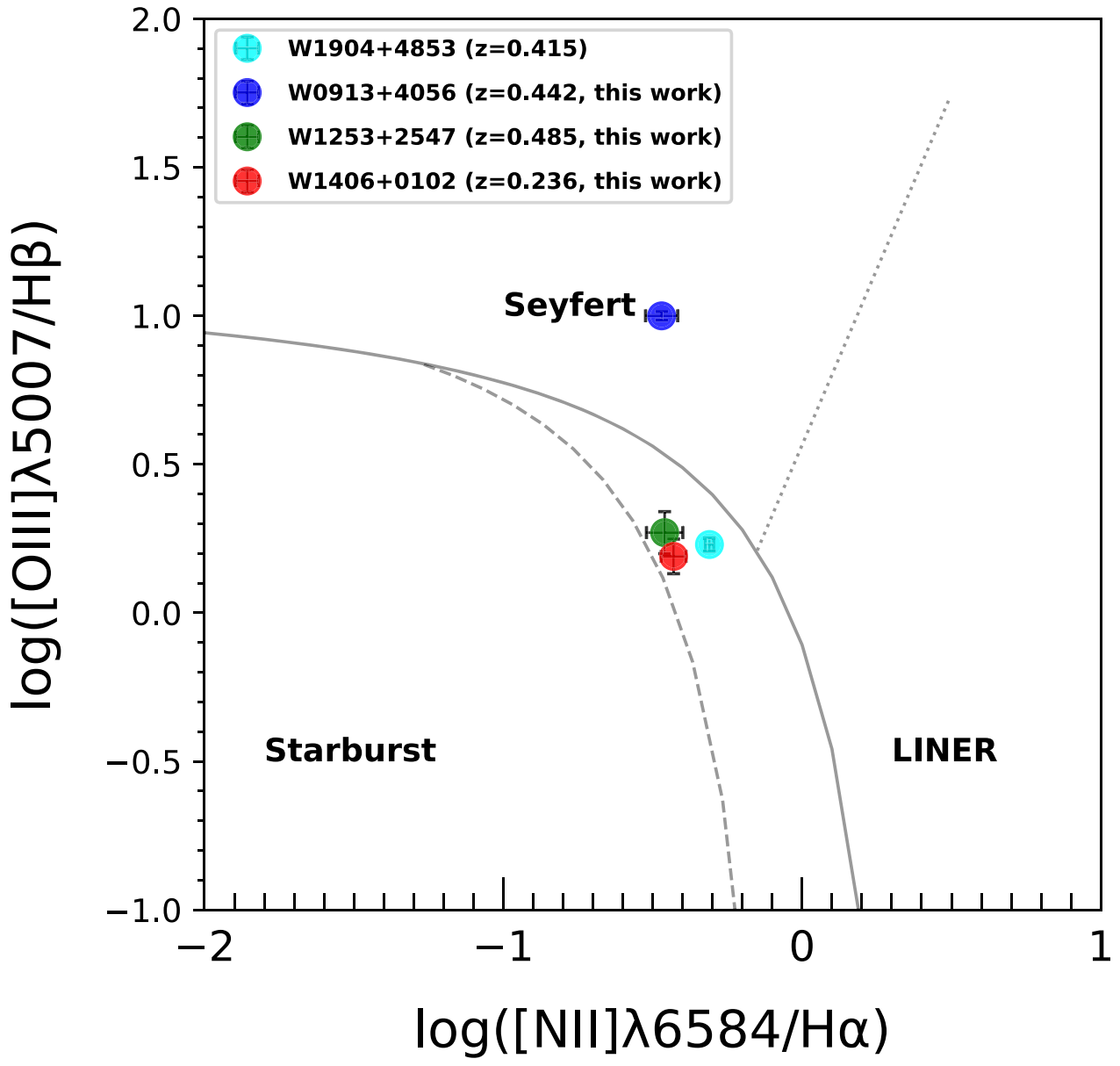

**Figure 6.** BPT diagram. The points with error bars represent the narrow-line fluxes of low-$z$ Hot DOGs. Solid lines show the demarcation between starburst galaxies and AGN defined by L. J. Kewley et al. (2001). The dashed line denotes the dividing line between star forming and composite star forming (G. Kauffmann et al. 2003). The dotted line separates Seyfert and LINER-type AGNs (L. J. Kewley et al. 2006).

### *4.4. The* $M_{BH}$–$M_{star}$ *Relation*

Following R. J. Assef et al. (2015) and J. Wu et al. (2018), G. Li et al. (2024) updated the relation between $M_{BH}$ and $M_{star}$ for high-$z$ ($z \sim$ 1–4) Hot DOGs, finding that the position of Hot DOGs is significantly above the local relation between the spheroidal component mass ($M_{sph}$) and $M_{BH}$ and overlaps with the region of $z \sim 6$ quasars. Figure 7 displays the $M_{BH}$–$M_{sph}$ relation for Hot DOGs in our sample, especially for the low-$z$ systems, drawing upon the $M_{BH}$ values from Section 4.2 and estimating $M_{star}$ as described in Section 3.5. We note that $M_{star}$ represents an upper limit for $M_{sph}$, as it includes the stellar mass from all components of the host galaxy, not just the spheroidal component.

Like their high-$z$ counterparts, these low-$z$ Hot DOGs also lie above the local $M_{BH}$–$M_{sph}$ relation, although they seem to be significantly closer to it. There are at least three potential scenarios that arise from this observation: (i) If the host galaxies of Hot DOGs ultimately end up in the local $M_{BH}$–$M_{sph}$ relation at $z = 0$, the offset with respect to the local relation may indicate that the quasar feedback during the Hot DOG stage is not sufficient to quench star formation. In this scenario, the fact that $z < 0.5$ Hot DOGs lie closer to the relation than high-$z$ ones would indicate a substantial difference in their star formation and nuclear accretion histories. Feedback from other processes—perhaps lower-luminosity AGN or radio-bright activity extending over timescales much longer than the ∼0.5 Myr lifetime of Hot DOGs proposed by G. Li et al. (2024)—might be more effective in regulating star formation. This might be consistent with the results from the SED modeling and BPT diagnostics, which indicate strong star formation within these systems despite the strong quasar activity. (ii) The proximity of low-$z$ Hot DOGs to the local $M_{\rm BH}$–$M_{\rm sph}$ may indicate that if the Hot DOG phase is a recurring event in these objects, as suggested by T. Dìaz-Santos et al. (2018), their cumulative effect of lowering the star formation efficiency and/or the gas mass reservoir of the host progressively with each occurrence may be sufficient over the age of the Universe to quench star formation. In this scenario, their host galaxies are not guaranteed to end up in the local $M_{\rm BH}$–$M_{\rm sph}$ relation at $z = 0$, and how close they get to it will depend on how many such cycles are left on average for $z < 0.5$ Hot DOGs. (iii) Hot DOGs may eventually become outliers from the local $M_{\rm BH}$–$M_{\rm sph}$ relation at $z = 0$. The fact that these low-$z$ Hot DOGs are closer to the local relation compared to the $z \sim 2$–4 systems might simply indicate the Hot DOG phase may be triggered at any point in the lifetime of the host galaxy and that star formation is being shut down at a later stage in the host's evolution, when a larger part of their initial gas reservoir has already been converted to stars.

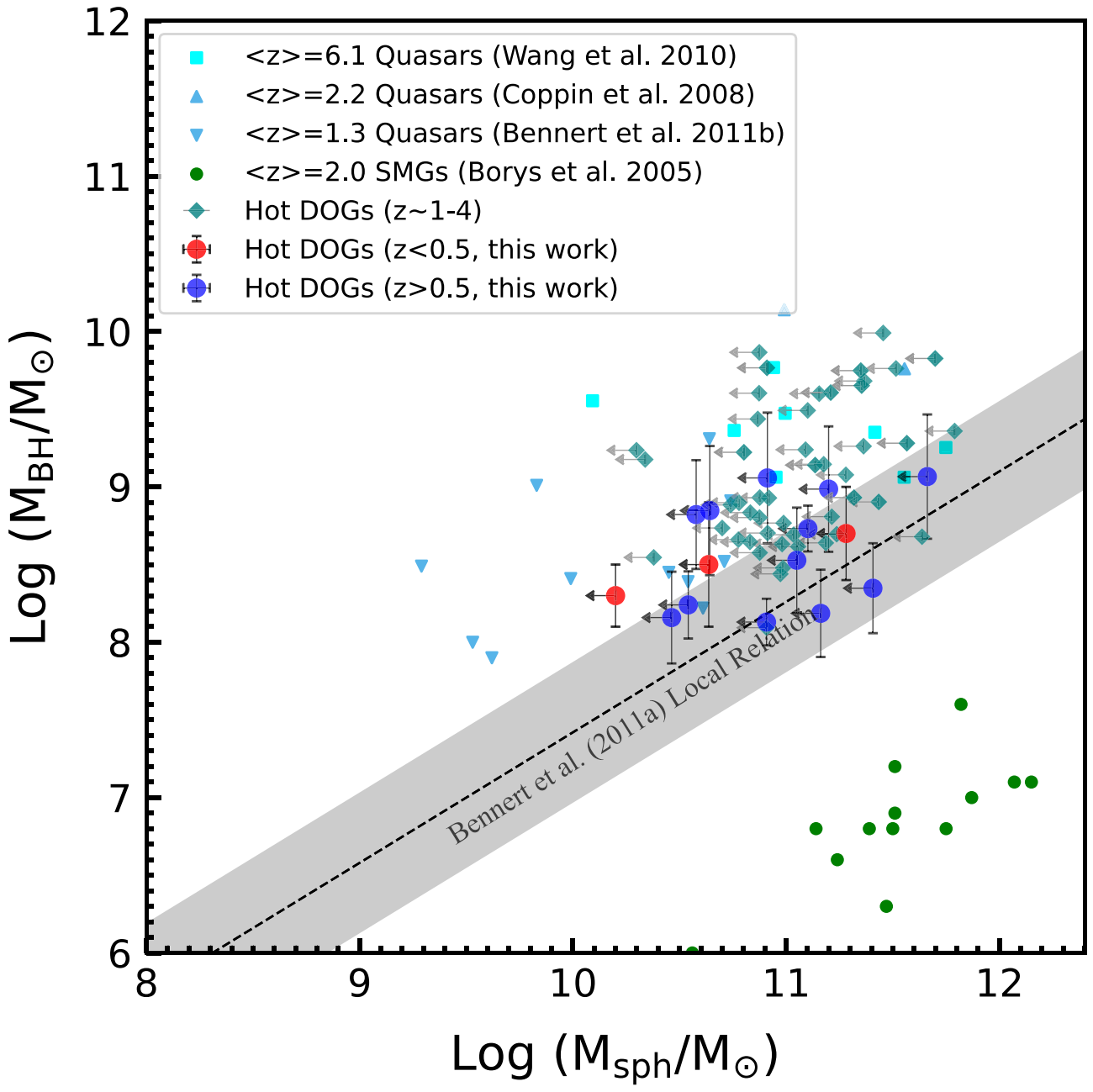

**Figure 7.** $M_{\rm BH}$ vs. $M_{\rm sph}$, following G. Li et al. (2024). The red points with error bars represent low-$z$ Hot DOGs. The bulge masses of host galaxies are predicted by the best-fit curve and subtracted from the potential AGN contribution, which is the upper limit. A local relation of active galaxies determined by V. N. Bennert et al. (2011), as well as data for SMG at $z \sim 2$ (C. Borys et al. 2005) and quasars at $z \sim 6$ (R. Wang et al. 2010), $z \sim 2$ (K. E. K. Coppin et al. 2008), and $z \sim 1.3$ (V. N. Bennert et al. 2011), are shown for comparison.

### *4.5. Number Density*

Considering the discontinuous nature of the Herschel survey, we recalculated the sky coverage of Herschel/PACS (~4.7%) and Herschel/SPIRE (~8.2%) using the corresponding field of view (G. Marton et al. 2017; B. Schulz et al. 2017), and found that the overlap between them in our sample is ~3%. The number of identified low-$z$ Hot DOGs (three) indicates that their surface number density is $N_{\rm HD} \sim 0.0024$ deg$^{-2}$, or approximately one for every ~400 deg$^2$, which is about an order of magnitude lower than those at high redshift (z ~ 2–4; $0.032 \pm 0.004$ deg$^{-2}$; R. J. Assef et al. 2015). This could be a consequence and tied to the overall evolution of the molecular gas density of galaxies across cosmic time (e.g., S. Berta et al. 2013; R. Decarli et al. 2016, 2019, 2020; D. A. Riechers et al. 2019, 2020; T. K. Garratt et al. 2021), characterized by peaking around redshift ~2 and decreasing steadily by approximately an order of magnitude to the present day.

To explore the relation between low-$z$ Hot DOGs and unobscured quasars, we use the quasar catalog from B. W. Lyke et al. (2020), which represents the largest quasar catalog obtained from SDSS DR16 (R. Ahumada et al. 2020). Within this catalog, there are 136 low-$z$ QSOs with $L_{\rm bol} > 10^{12.5} L_\odot$ and broad lines with FWHM > 1000 km s$^{-1}$ from 9376 deg$^2$ of the sky. Thus, the $N_{\rm HD}$ of low-$z$ unobscured quasars is 0.0145 deg$^{-2}$, 6 times that of low-$z$ Hot DOGs. However, there are six low-$z$ hyperluminous ($L_{\rm bol} > 10^{13} L_\odot$) objects in this catalog, suggesting that, within ~1200 deg$^2$, we would expect to find one unobscured hyperluminous quasar, consistent with the number of hyperluminous low-$z$ Hot DOGs. This could suggest a different shape of the luminosity function for $z < 0.5$ Hot DOGs than equally luminous type 1 quasars, but the low number statistics makes it difficult to draw a robust conclusion.

## 5. Conclusion

In this study, we utilized the WISE and Herschel archives to construct a sample of Hot DOGs across 3% of the sky, with a focus on low-$z$ ($z < 0.5$) candidates. Through SED fitting and spectral analysis, we identified three low-$z$ Hot DOGs. Our main results are summarized below.

1. Two of the spectroscopically confirmed low-$z$ systems, W1253+2547 and W1406+0102, show optical SEDs dominated by the host galaxy, while the other, W0913+4056, has an optical SED seemingly dominated by scattered light from the obscured central engine. This split is analogous to what is observed for the traditional high-$z$ Hot DOG population (R. J. Assef et al. 2016, 2020, 2022), and we refer to the host-dominated ones as "regular Hot DOGs" and the other as a blue excess Hot DOG or BHD.
2. The BPT diagnostics place the low-$z$ regular Hot DOGs in the composite area between AGN and star-forming galaxies while the BHD is located in the AGN region. We discuss that this could be evidence that BHDs are evolutionarily closer to the traditional type 1 AGN phase, as discussed in R. J. Assef et al. (2022) and G. Li et al. (2024).
3. The accretion of low-$z$ Hot DOGs approaches or is at the Eddington limit, and they also position above the local $M_{\rm BH}$–$M_{\rm sph}$ relation. If their host galaxies eventually fall into this relation by $z = 0$, it suggests that the quasar feedback during the Hot DOG phase is not sufficient to quench star formation. Otherwise, they may eventually become outliers from the local relation, which is the other scenario explored in G. Li et al. (2024). Compared to high-$z$ counterparts, these low-$z$ Hot DOGs seem to be significantly closer to the local $M_{\rm BH}$–$M_{\rm sph}$ relation, indicating that if the Hot DOG phase is a recurring event, its cumulative effect may be sufficient over the age of the Universe to quench star formation.
4. The surface number density of low-$z$ Hot DOGs is 0.0024 deg$^{-2}$, an order of magnitude lower than high-$z$ systems

estimated in R. J. Assef et al. (2015), consistent with the evolution of the cosmic gas density. Compared to hyperluminous unobscured AGNs in the same redshift range, the number of hyperluminous low-$z$ Hot DOGs is comparable. These results further support the interpretation that Hot DOGs may be a transitional phase in galaxy evolution.

## Acknowledgments

The systematic search and discovery of low-$z$ Hot DOGs compensate for their previous absence at low redshift, suggesting that these systems may represent the rapidly evolving "blowout" phase in galaxy evolution. These systems can serve as local analogs for studying highly luminous, highly obscured AGNs in the coevolution process of black holes and their host galaxies.

This work was supported by a grant from the National Natural Science Foundation of China (Nos. 11988101, 11973051, 12041302) and the International Partnership Program of Chinese Academy of Sciences, program No. 114A11KYSB20210010. G.D.L., C.-W.T., and R.J.A. are supported by the International Partnership Program of Chinese Academy of Sciences, program No. 114A11KYSB20210010. J.W. is thankful for support from the Tianchi Talent Program of Xinjiang Uygur Autonomous Region. R.J.A. was supported by FONDECYT grant No. 1231718 and by the ANID BASAL project FB210003. T.D.-S. acknowledges the research project was supported by the Hellenic Foundation for Research and Innovation (HFRI) under the "2nd Call for HFRI Research Projects to support Faculty Members & Researchers" (Project Number: 03382). G.D.L. would like to express thanks for the support from the National Natural Science Foundation of China Grant No. 11991052.

This publication makes use of data products from the Wide-field Infrared Survey Explorer, which is a joint project of the UCLA and the JPL/Caltech funded by the NASA, and from NEOWISE, which is a JPL/ Caltech project funded by NASA. Portions of this research were carried out at the Jet Propulsion Laboratory, California Institute of Technology, under a contract with the National Aeronautics and Space Administration. This work is also based on observations made with Herschel, a European Space Agency Cornerstone Mission with significant participation by NASA. Support for this work was provided by NASA through an award issued by JPL/Caltech.

Some of the data are based on observations obtained at the Hale Telescope, Palomar Observatory as part of a continuing collaboration between the California Institute of Technology, NASA/JPL, and Cornell University. Some of the data are based on the public science archive, including the Pan-STARRS1 DR2 Catalog at doi:10.17909/s0zg-jx37 and the Legacy Surveys.

*Facilities*: Palomar, DESI, GALEX, Pan-STARRS, SDSS, UKIRT, 2MASS, WISE, Herschel.

## Appendix A
## Physical Properties of High-$z$ Hot DOGs and IRGs

In this appendix, we provide the details and properties of the identified Hot DOGs (Table A1) and red galaxies (Table A2) in our sample. The criteria for object identification are described in Section 2.1, while the calculations of their properties are outlined in Section 3.

**Table A1**
Identified High-$z$ Hot DOGs with Spectroscopic Redshifts

| ID | R.A. (deg) | Decl. (deg) | $z_{\rm phot}$ | $z_{\rm spec}$ | log $L_{bol}$ ($L_\odot$) | $T_{\rm Max}$ (K) | $T_{\rm Min}$ (K) | log $M_{\rm dust}$ ($M_\odot$) | log $M_{\rm star}$ ($M_\odot$) | log $M_{\rm BH}$ ($M_\odot$) | log $\eta_{\rm Edd}$ |
|---|---|---|---|---|---|---|---|---|---|---|---|
| WISE J033114.48−275546.5 | 52.8103341 | −27.929594 | 1.036 | 1.591 | 13.0 | 900 ± 40 | 40 ± 4 | 7.9 | 10.4 | | |
| WISE J033532.17−284800.6 | 53.8840804 | −28.8001682 | 0.65 | 0.807 | 12.7 | 700 ± 10 | 28 ± 2 | 8.0 | 11.1 | | |
| WISE J085817.01+432643.2 | 134.5708944 | 43.4453564 | 0.953 | 2.647 | 13.8 | 1110 ± 50 | 49 ± 5 | 7.7 | 11.3 | | |
| WISE J085950.95−003756.3 | 134.962333 | −0.632323 | 1.14 | 1.887 | 13.7 | 1100 ± 20 | 36 ± 3 | 8.3 | 11.0 | | |
| WISE J094524.16−084007.6 | 146.3506869 | −8.6687953 | 0.516 | 0.692 | 13.0 | 1300 ± 30 | 63 ± 3 | 6.6 | 10.5 | 8.2 ± 0.3 (Mg II) | 0.3 ± 0.3 |
| WISE J100552.62+493448.0 | 151.4692854 | 49.5800165 | 0.402 | 1.122 | 13.8 | 1010 ± 10 | 83 ± 2 | 6.8 | | 9.2 ± 0.4 (Mg II) | 0.1 ± 0.4 |
| WISE J103803.36+572701.6 | 159.514001 | 57.4504552 | 0.784 | 1.284 | 13.3 | 1100 ± 10 | 33 ± 1 | 8.2 | 11.3 | 8.9 ± 0.4 (Mg II) | −0.1 ± 0.4 |
| WISE J104409.99+585225.0 | 161.0416327 | 58.8736272 | 0.962 | 2.54 | 13.5 | 1360 ± 20 | 53 ± 5 | 7.0 | 10.7 | | |
| WISE J121906.56+160243.1 | 184.7773716 | 16.0453065 | 0.759 | 0.761 | 12.7 | 1200 ± 100 | 36 ± 3 | 7.4 | 11.1 | 8.5 ± 0.3 (Mg II) | −0.3 ± 0.3 |
| WISE J121924.14+164158.8 | 184.8505863 | 16.6996928 | 0.633 | 1.105 | 12.9 | 1290 ± 10 | 38 ± 6 | 7.6 | 11 | 8.8 ± 0.4 (Mg II) | −0.3 ± 0.4 |
| WISE J124659.83+310754.2 | 191.7493138 | 31.1317459 | 0.47 | 0.564 | 12.7 | 1000 ± 30 | 34 ± 1 | 7.7 | 10.9 | 8.1 ± 0.2 (Hβ) | 0.1 ± 0.3 |
| WISE J124915.41+324357.2 | 192.3142251 | 32.732575 | 0.404 | 0.679 | 13.1 | 1260 ± 10 | 28 ± 1 | 8.2 | 11.2 | 8.2 ± 0.3 (Mg II) | 0.3 ± 0.3 |
| WISE J125522.89+271503.6 | 193.8453913 | 27.2510228 | 0.748 | 2.738 | 13.7 | 1250 ± 50 | 59 ± 5 | 7.4 | 11.7 | 9.1 ± 0.3 (C IV) | 0.1 ± 0.3 |
| WISE J125625.39+234804.4 | 194.1057934 | 23.8012408 | 0.743 | 0.885 | 13.2 | 1160 ± 30 | 48 ± 2 | 7.4 | 10.6 | 8.8 ± 0.4 (Mg II) | −0.2 ± 0.4 |
| WISE J125925.37+253606.8 | 194.8557115 | 25.601902 | 1.064 | 0.922 | 12.7 | 1020 ± 30 | 27 ± 4 | 8 | 11 | | |
| WISE J132609.90−113107.3 | 201.5412802 | −11.5187002 | | 1.335 | 13.3 | 1160 ± 50 | 26 ± 2 | 8.7 | 10.9 | 9.1 ± 0.4 (Mg II) | −0.4 ± 0.4 |
| WISE J133332.09+503519.6 | 203.3837168 | 50.5887918 | 0.635 | 0.801 | 12.9 | 930 ± 20 | 25 ± 1 | 8.3 | 11.2 | | |
| WISE J133739.92+000011.1 | 204.4163392 | 0.0030891 | 1.189 | 2.319 | 13.7 | 670 ± 60 | 38 ± 5 | 8.2 | 11.1 | | |
| WISE J142852.18+345159.9 | 217.2174326 | 34.8666575 | 0.797 | 1.037 | 13.1 | 830 ± 30 | 36 ± 3 | 8 | 11.4 | 8.3 ± 0.3 (Mg II) | 0.3 ± 0.3 |
| WISE J143051.89+335546.6 | 217.7162471 | 33.9296113 | 1.247 | 2.618 | 13.6 | 900 ± 100 | 43 ± 5 | 7.9 | 11.1 | 8.7 ± 0.3 (C IV) | 0.4 ± 0.3 |
| WISE J144309.50+013309.0 | 220.7896097 | 1.5525106 | 0.492 | 0.671 | 12.9 | 1260 ± 9 | 25 ± 1 | 8.3 | 10.5 | 8.2 ± 0.3 (Hβ) | 0.1 ± 0.3 |
| WISE J144526.92+174001.7 | 221.362185 | 17.6671451 | 1.171 | 1.982 | 13.9 | 1270 ± 20 | 42 ± 1 | 8.2 | 11.1 | | |
| WISE J161652.03+554607.6 | 244.2168277 | 55.7687999 | 1.154 | 1.173 | 12.9 | 970 ± 20 | 32 ± 2 | 7.9 | 10.5 | | |
| WISE J164250.09+412318.1 | 250.708742 | 41.3883871 | 1.002 | 1.276 | 13.1 | 970 ± 40 | 36 ± 7 | 7.8 | 11.2 | 9.0 ± 0.4 (Mg II) | −0.4 ± 0.4 |
| WISE J170535.38+603456.0 | 256.3974538 | 60.5822279 | 0.976 | 1.708 | 13.7 | 1400 ± 40 | 29 ± 2 | 8.8 | | 8.8 ± 0.3 (Mg II) | 0.4 ± 0.3 |
| WISE J200511.41+021517.6 | 301.2975498 | 2.2548958 | | 0.782 | 12.5 | 690 ± 20 | 19 ± 3 | 8.6 | 10.5 | | |

**Note.** Spectroscopic redshifts are from our Palomar/DBSP observations, SDSS (Abdurro'uf et al. 2022), DESI (DESI Collaboration et al. 2024a, 2024b), and the literature (J. D. Silverman et al. 2010; E. Kalfountzou et al. 2011; A. Hernán-Caballero & E. Hatziminaoglou 2011; I. K. Baldry et al. 2018; P. R. M. Eisenhardt et al. 2025, in preparation). The dust temperature and dust mass are estimated using the SPL model, as described in Section 3.2. For sources where Hα, Hβ, Mg II, or C IV are detected, we calculate $M_{\rm BH}$ as described in Section 3.6.

**Table A2**
The Properties of Selected Red Galaxies

| ID | R.A. (deg) | Decl. (deg) | $z_{phot}$ | $z_{spec}$ | $\log L_{bol}$ ($L_\odot$) | $\log M_{star}$ ($M_\odot$) | $\log M_{BH}$ ($M_\odot$) | $\log \eta_{Edd}$ |
|---|---|---|---|---|---|---|---|---|
| WISE J005057.25+432038.3 | 12.7385461 | 43.3439942 | ⋯ | 0.187 | 11.9 | 10.4 | 8.1 ± 0.3 (Hα) | −0.7 ± 0.3 |
| WISE J012002.63+142142.4 | 20.0109749 | 14.3617917 | 0.022 | 0.031 | 11.6 | 10.2 | ⋯ | ⋯ |
| WISE J015238.58+005942.4 | 28.1607762 | 0.9951208 | 0.283 | 0.229 | 11.3 | 9.8 | ⋯ | ⋯ |
| WISE J031501.43+420208.8 | 48.7559668 | 42.0357953 | ⋯ | 0.023 | 11.3 | 9.6 | 7.4 ± 0.2 (Hα) | −0.6 ± 0.2 |
| WISE J033620.45−272422.2 | 54.0852417 | −27.4061811 | 0.367 | 0.346 | 12.1 | 10.6 | ⋯ | ⋯ |
| WISE J084116.60+020014.7 | 130.3191759 | 2.0041087 | 0.516 | 0.388 | 12.3 | 10.8 | ⋯ | ⋯ |
| WISE J090414.10−002144.8 | 136.0587727 | −0.3624675 | 0.269 | 0.353 | 12.1 | 10.6 | ⋯ | ⋯ |
| WISE J091237.75−005506.5 | 138.1573085 | −0.9184892 | 0.312 | 0.256 | 11.8 | 10.2 | ⋯ | ⋯ |
| WISE J095856.19+024128.1 | 149.7341268 | 2.691153 | 0.685 | 0.91 | 12.6 | 10.6 | ⋯ | ⋯ |
| WISE J095943.61+014407.4 | 149.9317289 | 1.7353982 | ⋯ | 1.427 | 12.8 | 10.8 | ⋯ | ⋯ |
| WISE J110151.93+571601.8 | 165.4664116 | 57.2671921 | 0.401 | 0.354 | 12.1 | 10.7 | 8.0 ± 0.3 (Hβ) | −0.4 ± 0.3 |
| WISE J110952.83+423315.7 | 167.4701657 | 42.5543651 | 0.162 | 0.261 | 12.3 | 10.6 | 8.1 ± 0.2 (Hα) | −0.3 ± 0.2 |
| WISE J125059.31+264244.9 | 192.7471536 | 26.7124878 | 0.212 | 0.187 | 11.8 | 9.9 | ⋯ | ⋯ |
| WISE J125527.48+252423.7 | 193.8645153 | 25.4066096 | 0.282 | 0.282 | 11.6 | 10.4 | ⋯ | ⋯ |
| WISE J130247.18+293157.3 | 195.6966226 | 29.5325996 | 0.450 | 0.392 | 12.2 | 10.4 | 8.1 ± 0.2 (Hα) | −0.4 ± 0.2 |
| WISE J130813.37+330333.8 | 197.0557205 | 33.0594043 | 0.668 | 0.668 | 12.6 | 11.1 | ⋯ | ⋯ |
| WISE J131256.60+240310.1 | 198.2358349 | 24.0528165 | 0.278 | 0.627 | 12.5 | 10.8 | ⋯ | ⋯ |
| WISE J131421.26+345245.4 | 198.5886066 | 34.8792826 | 0.218 | 0.238 | 12.0 | 10.9 | ⋯ | ⋯ |
| WISE J131654.02+234046.4 | 199.2250982 | 23.6795686 | 0.150 | 0.138 | 11.8 | 10.4 | ⋯ | ⋯ |
| WISE J133303.66+502825.2 | 203.2652514 | 50.4736687 | 0.377 | 0.313 | 11.7 | 10.4 | ⋯ | ⋯ |
| WISE J133926.07+253519.8 | 204.8586427 | 25.5888492 | 0.112 | 0.076 | 11.2 | 10.0 | 7.4 ± 0.2 (Hα) | −0.7 ± 0.2 |
| WISE J134442.09+555313.2 | 206.1754092 | 55.8870195 | 0.028 | 0.037 | 12.1 | 10.6 | 7.9 ± 0.2 (Hα) | −0.4 ± 0.2 |
| WISE J134617.74+300107.2 | 206.5739542 | 30.0186705 | 0.492 | 0.410 | 12.0 | 10.3 | 7.5 ± 0.2 (Hα) | −0.6 ± 0.2 |
| WISE J134705.55+281805.3 | 206.7731649 | 28.3014797 | 0.241 | 0.255 | 12.4 | 10.7 | ⋯ | ⋯ |
| WISE J134706.92+345624.0 | 206.7788607 | 34.9400256 | 0.044 | 0.054 | 11.3 | 9.4 | ⋯ | ⋯ |
| WISE J140912.46−013455.2 | 212.3019458 | −1.5820129 | 0.273 | 0.265 | 11.8 | 10.7 | ⋯ | ⋯ |
| WISE J161300.87+534131.1 | 243.2536563 | 53.6919877 | 0.937 | 0.886 | 12.6 | 10.7 | ⋯ | ⋯ |
| WISE J164021.03+462843.2 | 250.0876266 | 46.4786866 | 0.025 | 0.059 | 11.0 | 9.2 | 7.1 ± 0.2 (Hα) | −0.6 ± 0.2 |
| WISE J171144.19+585226.7 | 257.9341538 | 58.8740864 | 1.127 | 1.189 | 12.8 | 10.7 | ⋯ | ⋯ |
| WISE J205826.80−423900.4 | 314.611687 | −42.6501198 | 0.031 | 0.043 | 12.0 | 10.5 | ⋯ | ⋯ |
| WISE J221159.98−114535.4 | 332.9999216 | −11.7598478 | 0.240 | 0.362 | 12.2 | 10.7 | 7.9 ± 0.3 (Mg II) | −0.2 ± 0.3 |
| WISE J231546.76−590314.6 | 348.9448461 | −59.0540763 | 0.025 | 0.045 | 12.0 | 10.5 | ⋯ | ⋯ |

**Note.** Spectroscopic redshift obtained from our Palomar/DBSP observations, SDSS (Abdurro'uf et al. 2022), and DESI (DESI Collaboration et al. 2024a, 2024b). We estimate $M_{star}$, $L_{bol}$, and $M_{BH}$ as described in Section 3.

# Appendix B
# Spectral Observations of High-$z$ Hot DOGs and Infrared Galaxies

We present the optical spectra of Hot DOGs (Figure B1) and red galaxies (Figure B2) observed with the Palomar/DBSP. The corresponding spectral line measurements are listed in Table B1.

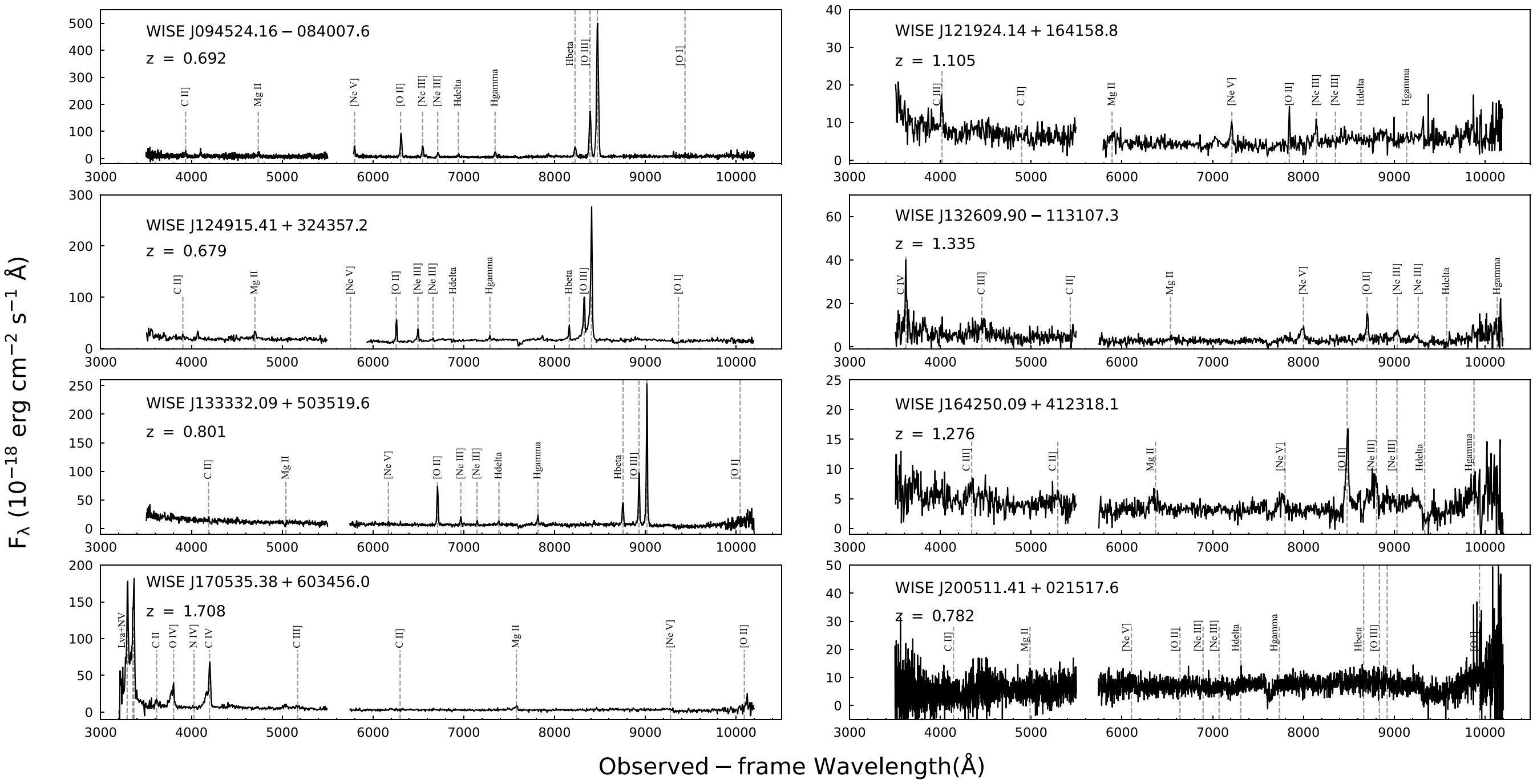

**Figure B1.** Optical spectra of Hot DOG candidates obtained at Palomar Observatory. The spectra cover wavelengths 3200–10,500 Å in the observed frame, with a gap from 5500 Å to 5,700 Å due to the dichroic mirror.

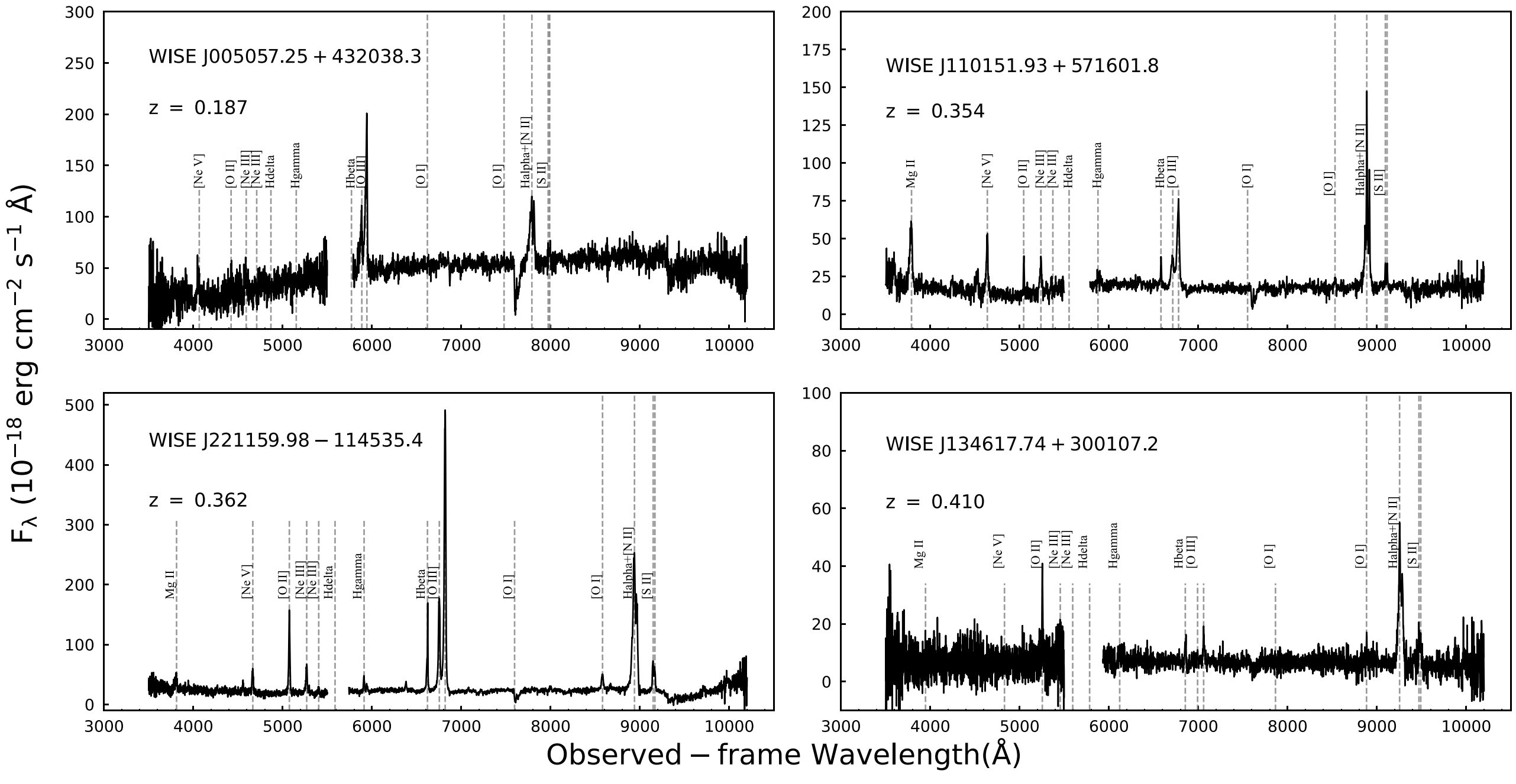

**Figure B2.** Optical spectra of IRGs outside of our sample obtained at Palomar Observatory. See Section 2.2.

**Table B1**
Properties of Emission Lines for Galaxies Observed with Palomar/DBSP

| Name | Line | log $L$ (erg s$^{-1}$) | log $L_{\rm bl}$ (erg s$^{-1}$) | $\triangle v_{\rm bl}$ (km s$^{-1}$) | FWHM$_{\rm bl}$ (km s$^{-1}$) | EW$_{\rm bl}$ (ergs$^{-1}$) | log $L_{\rm nl}$ (Å) | $\triangle v_{\rm nl}$ (km s$^{-1}$) | FWHM$_{\rm nl}$ (km s$^{-1}$) | EW$_{\rm nl}$ (Å) |
|---|---|---|---|---|---|---|---|---|---|---|
| W0945-0840 | Mg II | 41.54 ± 0.07 | 41.54 ± 0.07 | −180 ± 90 | 2820 ± 20 | 9 ± 1 | … | … | … | … |
| | [O II] | 42.54 ± 0.01 | … | … | … | … | 42.54 ± 0.01 | 16 ± 5 | 820 ± 10 | 130 ± 8 |
| | [Ne III]λ3868 | 42.29 ± 0.02 | … | … | … | … | 42.29 ± 0.02 | 50 ± 10 | 930 ± 20 | 155 ± 6 |
| | Hζ | 41.60 ± 0.03 | … | … | … | … | 41.60 ± 0.03 | 140 ± 30 | 780 ± 40 | 32 ± 2 |
| | [Ne III]λ3967 | 41.98 ± 0.02 | … | … | … | … | 41.98 ± 0.02 | 170 ± 10 | 930 ± 20 | 79 ± 4 |
| | Hδ | 41.57 ± 0.03 | … | … | … | … | 41.57 ± 0.03 | −20 ± 20 | 720 ± 60 | 20 ± 1 |
| | Hγ | 42.04 ± 0.02 | … | … | … | … | 42.04 ± 0.02 | 90 ± 20 | 880 ± 30 | 82 ± 5 |
| | [O III]λ4363 | 41.55 ± 0.03 | … | … | … | … | 41.55 ± 0.03 | 150 ± 20 | 470 ± 30 | 27 ± 2 |
| | Hβ | 42.29 ± 0.01 | … | … | … | … | 42.29 ± 0.01 | 80 ± 10 | 880 ± 30 | 74 ± 5 |
| | [O III]λ4959 | 42.91 ± 0.01 | … | … | … | … | 42.91 ± 0.01 | 63 ± 2 | 794 ± 7 | 190 ± 10 |
| | [O III]λ5007 | 43.39 ± 0.01 | … | … | … | … | 43.39 ± 0.01 | 57 ± 1 | 799 ± 2 | 440 ± 10 |
| W1101+5716 | Mg II | 41.80 ± 0.01 | 41.80 ± 0.01 | −70 ± 20 | 2740 ± 60 | 60 ± 2 | … | … | … | … |
| | [O II] | 41.02 ± 0.02 | … | … | … | … | 41.02 ± 0.02 | −110 ± 10 | 550 ± 30 | 13 ± 1 |
| | [Ne III]λ3868 | 41.37 ± 0.03 | 41.3 ± 0.1 | −170 ± 80 | 1800 ± 200 | 22 ± 5 | 40.7 ± 0.2 | 90 ± 50 | 700 ± 200 | 7 ± 3 |
| | Hβ | 41.05 ± 0.07 | 40.6 ± 0.2 | −300 ± 200 | 1800 ± 500 | 4 ± 2 | 40.84 ± 0.05 | 20 ± 10 | 380 ± 40 | 6 ± 1 |
| | [O III]λ4959 | 41.66 ± 0.02 | 41.66 ± 0.02 | 3 ± 30 | 2300 ± 100 | 47 ± 2 | … | … | … | … |
| | [O III]λ5007 | 41.89 ± 0.01 | 41.89 ± 0.01 | 170 ± 10 | 1380 ± 30 | 80 ± 2 | … | … | … | … |
| | [O I]λ6300 | 40.78 ± 0.09 | … | … | … | … | 40.78 ± 0.09 | 50 ± 60 | 700 ± 100 | 6 ± 1 |
| | Hα | 42.09 ± 0.02 | 41.95 ± 0.04 | 70 ± 40 | 2000 ± 100 | 90 ± 10 | 41.55 ± 0.07 | 20 ± 10 | 250 ± 30 | 36 ± 5 |
| | [N II]λ6548 | 40.8 ± 0.1 | … | … | … | … | 40.8 ± 0.1 | 20 ± 20 | 260 ± 30 | 7 ± 2 |
| | [N II]λ6584 | 41.34 ± 0.08 | … | … | … | … | 41.34 ± 0.08 | 20 ± 10 | 260 ± 30 | 22 ± 3 |
| | [S II]λ6716 | 40.76 ± 0.06 | … | … | … | … | 40.76 ± 0.06 | 20 ± 10 | 310 ± 30 | 5 ± 1 |
| | [S II]λ6731 | 40.74 ± 0.06 | … | … | … | … | 40.74 ± 0.06 | 20 ± 10 | 290 ± 40 | 5 ± 1 |
| W1219+1641 | C III]] | 42.06 ± 0.03 | 42.06 ± 0.03 | −330 ± 50 | 1400 ± 100 | 9 ± 1 | … | … | … | … |
| | Mg II | 42.0 ± 0.1 | 42.0 ± 0.1 | −60 ± 200 | 3400 ± 900 | 14 ± 4 | … | … | … | … |
| | [Ne V] λ3426 | 42.06 ± 0.05 | 42.06 ± 0.05 | −190 ± 60 | 1100 ± 150 | 19 ± 1 | … | … | … | … |
| | [O II] | 41.99 ± 0.02 | … | … | … | … | 41.99 ± 0.02 | −200 ± 10 | 500 ± 30 | 16 ± 1 |
| W1249+3243 | Mg II | 41.79 ± −0.08 | 41.79 ± −0.08 | 110 ± 40 | 1170 ± 80 | 8 ± 2 | … | … | … | … |
| | [O II] | 42.15 ± 0.05 | … | … | … | … | 42.15 ± 0.05 | 124 ± 9 | 620 ± 20 | 31 ± 4 |
| | [Ne III]λ3868 | 42.07 ± 0.04 | 41.76 ± 0.09 | −430 ± 140 | 1960 ± 310 | 12 ± 3 | 41.78 ± 0.07 | 50 ± 30 | 530 ± 60 | 12 ± 3 |
| | Hγ | 41.5 ± 0.1 | … | … | … | … | 41.5 ± 0.1 | −10 ± 50 | 780 ± 220 | 6 ± 2 |
| | Hβ | 42.05 ± 0.04 | 41.5 ± 0.2 | -1200 ± 200 | 1460 ± 350 | 6 ± 2 | 41.89 ± 0.08 | −400 ± 10 | 460 ± 30 | 13 ± 2 |
| | [O III]λ4959 | 42.56 ± 0.02 | 42.12 ± −0.03 | -360 ± 100 | 1040 ± 160 | 16 ± 1 | 42.35 ± 0.02 | 69 ± 6 | 480 ± 20 | 27 ± 2 |
| | [O III]λ5007 | 43.03 ± 0.01 | 42.66 ± 0.01 | -370 ± 20 | 1070 ± 30 | 41 ± 1 | 42.78 ± 0.01 | 68 ± 2 | 450 ± 7 | 54 ± 2 |
| W1326-1131 | C IV | 42.80 ± 0.01 | 42.80 ± 0.01 | −60 ± 10 | 1560 ± 40 | 31 ± 1 | … | … | … | … |
| | Mg II | 42.14 ± 0.08 | 42.14 ± 0.08 | 600 ± 300 | 3900 ± 800 | 21 ± 4 | … | … | … | … |
| | [O II] | 42.54 ± 0.02 | 42.0 ± 0.1 | −600 ± 200 | 1000 ± 200 | 10 ± 3 | 42.41 ± 0.05 | −80 ± 20 | 660 ± 40 | 27 ± 4 |
| W1333+5035 | [O II] | 42.53 ± 0.01 | … | … | … | … | 42.53 ± 0.01 | −192 ± 1 | 698 ± 3 | 90 ± 1 |
| | [Ne III]λ3868 | 41.89 ± 0.02 | … | … | … | … | 41.89 ± 0.02 | −30 ± 20 | 670 ± 30 | 27 ± 2 |
| | Hζ | 41.34 ± 0.05 | … | … | … | … | 41.34 ± 0.05 | 30 ± 30 | 580 ± 50 | 8 ± 1 |
| | Hδ | 41.64 ± 0.04 | … | … | … | … | 41.64 ± 0.04 | −200 ± 40 | 900 ± 100 | 14 ± 1 |
| | Hγ | 41.87 ± 0.03 | … | … | … | … | 41.87 ± 0.03 | −70 ± 20 | 580 ± 30 | 19 ± 1 |
| | Hβ | 42.35 ± 0.01 | … | … | … | … | 42.35 ± 0.01 | −81 ± 2 | 580 ± 4 | 53 ± 1 |
| | [O III]λ4959 | 42.60 ± 0.01 | … | … | … | … | 42.60 ± 0.01 | 9.0 ± 0.6 | 457 ± 2 | 66 ± 1 |
| | [O III]λ5007 | 43.02 ± 0.01 | … | … | … | … | 43.02 ± 0.01 | 10.1 ± 0.2 | 447 ± 1 | 160 ± 1 |
| W1346+3001 | [O II] | 41.32 ± 0.04 | … | … | … | … | 41.32 ± 0.04 | 40 ± 10 | 620 ± 30 | 30 ± 3 |
| | Hβ | 40.9 ± 0.1 | … | … | … | … | 40.9 ± 0.1 | −370 ± 50 | 740 ± 100 | 15 ± 4 |
| | [O III]λ4959 | 40.5 ± 0.3 | … | … | … | … | 40.5 ± 0.3 | 10 ± 100 | 730 ± 220 | 5 ± 4 |
| | [O III]λ5007 | 41.0 ± 0.1 | … | … | … | … | 41.0 ± 0.1 | 50 ± 40 | 660 ± 120 | 14 ± 3 |

**Table B1**
(Continued)

| Name | Line | log $L$ (erg s$^{-1}$) | log $L_{bl}$ (erg s$^{-1}$) | $\Delta v_{bl}$ (km s$^{-1}$) | FWHM$_{bl}$ (km s$^{-1}$) | EW$_{bl}$ (ergs$^{-1}$) | log$L_{nl}$ (Å) | $\Delta v_{nl}$ (km s$^{-1}$) | FWHM$_{nl}$ (km s$^{-1}$) | EW$_{nl}$ (Å) |
|---|---|---|---|---|---|---|---|---|---|---|
| | Hα | 42.04 ± 0.04 | 41.97 ± 0.02 | 150 ± 50 | 1750 ± 110 | 100 ± 15 | 41.2 ± 0.1 | 10 ± 10 | 350 ± 40 | 36 ± 9 |
| | [N II]λ6548 | 40.2 ± 0.5 | ⋯ | ⋯ | ⋯ | ⋯ | 40.2 ± 0.5 | 80 ± 20 | 550 ± 100 | 3 ± 3 |
| | [N II]λ6584 | 41.25 ± 0.09 | ⋯ | ⋯ | ⋯ | ⋯ | 41.25 ± 0.09 | 80 ± 10 | 550 ± 100 | 38 ± 9 |
| | [S II]λ6716 | 40.93 ± 0.09 | ⋯ | ⋯ | ⋯ | ⋯ | 40.93 ± 0.09 | 20 ± 20 | 340 ± 50 | 14 ± 3 |
| | [S II]λ6731 | 40.9 ± 0.1 | ⋯ | ⋯ | ⋯ | ⋯ | 40.9 ± 0.1 | −30 ± 30 | 470 ± 90 | 12 ± 3 |
| W1642+4123 | C III]λ1909 | 42.08 ± 0.04 | 42.08 ± 0.04 | −300 ± 100 | 3300 ± 300 | 11 ± 1 | ⋯ | ⋯ | ⋯ | ⋯ |
| | Mg II | 42.2 ± 0.2 | 42.2 ± 0.2 | −1300 ± 500 | 3800 ± 500 | 23 ± 9 | ⋯ | ⋯ | ⋯ | ⋯ |
| | [O II] | 42.59 ± 0.02 | 42.59 ± 0.02 | −110 ± 10 | 1100 ± 40 | 42 ± 3 | ⋯ | ⋯ | ⋯ | ⋯ |
| W1705+6034 | Lyα | 44.06 ± 0.01 | 43.97 ± 0.01 | 383 ± 7 | 4850 ± 30 | 110.2 ± 0.8 | 43.32 ± 0.01 | 398 ± 2 | 860 ± 10 | 24.3 ± 0.6 |
| | Ne V] | 44.01 ± 0.01 | 44.01 ± 0.01 | 286 ± 1 | 2986 ± 7 | 119 ± 1 | ⋯ | ⋯ | ⋯ | ⋯ |
| | C IV | 43.66 ± 0.01 | 43.37 ± 0.01 | −2060 ± 50 | 4200 ± 1200 | 65 ± 2 | 43.35 ± 0.01 | 180 ± 4 | 1360 ± 20 | 62 ± 2 |
| | Mg II | 42.7 ± 0.1 | 42.7 ± 0.1 | −270 ± 60 | 1900 ± 160 | 30 ± 7 | ⋯ | ⋯ | ⋯ | ⋯ |
| W2211-1145 | Mg II | 41.3 ± 0.1 | 41.3 ± 0.1 | −300 ± 40 | 1500 ± 200 | 12 ± 4 | ⋯ | ⋯ | ⋯ | ⋯ |
| | Ne V] λ3346 | 41.07 ± 0.05 | 41.07 ± 0.05 | −100 ± 50 | 1000 ± 100 | 9 ± 1 | ⋯ | ⋯ | ⋯ | ⋯ |
| | Ne V] λ3426 | 41.52 ± 0.02 | ⋯ | ⋯ | ⋯ | ⋯ | 41.52 ± 0.02 | −140 ± 20 | 980 ± 50 | 24 ± 1 |
| | [O II] | 41.95 ± 0.01 | ⋯ | ⋯ | ⋯ | ⋯ | 41.95 ± 0.01 | −251 ± 4 | 824 ± 9 | 68 ± 2 |
| | [Ne III]λ3868 | 41.65 ± 0.02 | 41.65 ± 0.02 | −120 ± 20 | 1090 ± 60 | 36 ± 2 | ⋯ | ⋯ | ⋯ | ⋯ |
| | Hζ | 40.74 ± 0.07 | ⋯ | ⋯ | ⋯ | ⋯ | 40.74 ± 0.07 | −30 ± 40 | 530 ± 80 | 4.4 ± 0.7 |
| | [Ne III]λ3967 | 40.80 ± 0.07 | ⋯ | ⋯ | ⋯ | ⋯ | 40.80 ± 0.07 | −30 ± 50 | 700 ± 100 | 5.1 ± 0.8 |
| | Hγ | 41.28 ± 0.04 | ⋯ | ⋯ | ⋯ | ⋯ | 41.28 ± 0.04 | −90 ± 30 | 730 ± 80 | 14 ± 2 |
| | Hβ | 41.81 ± 0.01 | 41.64 ± 0.02 | −130 ± 10 | 770 ± 20 | 28 ± 1 | 41.32 ± 0.02 | 100 ± 2 | 181 ± 6 | 13.6 ± 0.6 |
| | [O III]λ4363 | 42.19 ± 0.01 | ⋯ | ⋯ | ⋯ | ⋯ | 42.19 ± 0.01 | −88 ± 4 | 900 ± 10 | 95 ± 3 |
| | [O III]λ4959 | 42.70 ± 0.01 | 41.86 ± 0.04 | −900 ± 70 | 1500 ± 100 | 44 ± 5 | 42.63 ± 0.01 | −69 ± 2 | 824 ± 5 | 260 ± 6 |
| | [O I]λ6300 | 41.45 ± 0.02 | ⋯ | ⋯ | ⋯ | ⋯ | 41.45 ± 0.02 | −50 ± 20 | 850 ± 40 | 18 ± 1 |
| | Hα | 42.49 ± 0.02 | 41.82 ± 0.07 | −1400 ± 200 | 2600 ± 300 | 47 ± 8 | 42.38 ± 0.02 | −100 ± 20 | 700 ± 40 | 170 ± 10 |
| | [N II]λ6548 | 41.90 ± 0.05 | ⋯ | ⋯ | ⋯ | ⋯ | 41.90 ± 0.05 | −120 ± 20 | 750 ± 30 | 57 ± 6 |
| | [N II]λ6584 | 42.23 ± 0.03 | ⋯ | ⋯ | ⋯ | ⋯ | 42.23 ± 0.03 | −120 ± 20 | 750 ± 30 | 123 ± 8 |
| | [S II]λ6716 | 41.49 ± 0.03 | ⋯ | ⋯ | ⋯ | ⋯ | 41.49 ± 0.03 | −110 ± 10 | 450 ± 20 | 20 ± 2 |
| | [S II]λ6731 | 41.48 ± 0.03 | ⋯ | ⋯ | ⋯ | ⋯ | 41.48 ± 0.03 | −130 ± 10 | 540 ± 30 | 20 ± 2 |
| W2228+7504 | [O II] | 38.05 ± 0.02 | ⋯ | ⋯ | ⋯ | ⋯ | 38.05 ± 0.02 | −89 ± 8 | 500 ± 20 | 80 ± 10 |
| | Hδ | 37.23 ± 0.09 | ⋯ | ⋯ | ⋯ | ⋯ | 37.23 ± 0.09 | 40 ± 30 | 400 ± 100 | 12 ± 3 |
| | Hγ | 37.41 ± 0.08 | ⋯ | ⋯ | ⋯ | ⋯ | 37.41 ± 0.08 | 10 ± 20 | 300 ± 50 | 16 ± 5 |
| | Hβ | 37.92 ± 0.02 | ⋯ | ⋯ | ⋯ | ⋯ | 37.92 ± 0.02 | 22 ± 4 | 250 ± 10 | 63 ± 7 |
| | [O III]λ4959 | 37.63 ± 0.04 | ⋯ | ⋯ | ⋯ | ⋯ | 37.63 ± 0.04 | 24 ± 9 | 260 ± 20 | 40 ± 10 |
| | [O III]λ5007 | 38.09 ± 0.01 | ⋯ | ⋯ | ⋯ | ⋯ | 38.09 ± 0.01 | 30 ± 3 | 240 ± 7 | 82 ± 8 |
| | Hα | 38.96 ± 0.01 | ⋯ | ⋯ | ⋯ | ⋯ | 38.96 ± 0.01 | 101 ± 1 | 290 ± 1 | 530 ± 30 |
| | [N II]λ6548 | 37.56 ± 0.05 | ⋯ | ⋯ | ⋯ | ⋯ | 37.56 ± 0.05 | 80 ± 10 | 300 ± 30 | 21 ± 3 |
| | [N II]λ6584 | 38.07 ± 0.01 | ⋯ | ⋯ | ⋯ | ⋯ | 38.07 ± 0.01 | 79 ± 5 | 290 ± 10 | 69 ± 5 |

**Note.** $L$ is the luminosity derived from the best-fit model, and $\Delta v$ indicates the offset of the Gaussian center for each broad/narrow component. Subscripts "bl" and "nl" indicate broad line and narrow line, respectively.

## ORCID iDs

Guodong Li https://orcid.org/0000-0003-4007-5771
Jingwen Wu https://orcid.org/0000-0001-7808-3756
Chao-Wei Tsai https://orcid.org/0000-0002-9390-9672
Daniel Stern https://orcid.org/0000-0003-2686-9241
Roberto J. Assef https://orcid.org/0000-0002-9508-3667
Kevin McCarthy https://orcid.org/0000-0001-6857-018X
Hyunsung D. Jun https://orcid.org/0000-0003-1470-5901
Tanio Díaz-Santos https://orcid.org/0000-0003-0699-6083
Andrew W. Blain https://orcid.org/0000-0001-7489-5167
Trystan Lambert https://orcid.org/0000-0001-6263-0970
Dejene Zewdie https://orcid.org/0000-0003-4293-7507
Román Fernández Aranda https://orcid.org/0000-0002-7714-688X
Cuihuan Li https://orcid.org/0000-0001-6405-9675